\documentclass[sigconf]{acmart}
\usepackage{listings}
\usepackage{color}
\usepackage{multirow}
\usepackage{booktabs}
\usepackage{graphicx}
\usepackage[skins]{tcolorbox}
\usepackage{colortbl}
\usepackage{geometry}
\usepackage{amsmath}
\usepackage{bm}
\usepackage{setspace}
\usepackage{enumitem}
\usepackage{xspace}
\usepackage{hyperref}

\setenumerate[1]{itemsep=0pt,partopsep=0pt,parsep=\parskip,topsep=5pt}
\setitemize[1]{itemsep=0pt,partopsep=0pt,parsep=\parskip,topsep=5pt}
\setdescription{itemsep=0pt,partopsep=0pt,parsep=\parskip,topsep=5pt}
\definecolor{dkgreen}{rgb}{0,0.6,0}
\definecolor{gray}{rgb}{0.5,0.5,0.5}
\definecolor{mauve}{rgb}{0.58,0,0.82}
\definecolor{keywordcolor}{rgb}{0.8,0.1,0.5}
\definecolor{webgreen}{rgb}{0,.5,0}
\acmConference[ICSE 2024]{46th International Conference on Software Engineering}{April 2024}{Lisbon, Portugal}

\tcbset{
  top=1mm,bottom=1pt,left=1pt,right=1pt,
}

\newcommand{\toolname}{\textsc{MicroFuzz}\xspace}
\newcommand{\company}{\textsc{AntGroup}\xspace}

\AtBeginDocument{%
  }

\title{\toolname: An Efficient Fuzzing Framework for Microservices}

\author{Peng Di}
\affiliation{%
  \institution{Ant Group}
  \city{Hangzhou}
  \country{China}
  }
\email{dipeng.dp@antgroup.com}

\author{Bingchang Liu}
\affiliation{%
  \institution{Ant Group}
  \city{Beijing}
  \country{China}
  }
\email{bingchang.lbc@antgroup.com}

\author{Yiyi Gao}
\affiliation{%
  \institution{Ant Group}
  \city{Hangzhou}
  \country{China}
  }
\email{gaoyiyi.gyy@antgroup.com}

\begin{abstract}

Fuzzing is a widely adopted technique in the software industry to enhance security and software quality. However, most existing fuzzers are specifically designed for monolithic software architectures and face significant limitations when it comes to serving distributed Microservices applications (Apps). These limitations primarily revolve around issues of inconsistency, communication, and applicability which arise due to the differences in monolithic and distributed software architecture.

This paper presents a novel fuzzing framework, called \toolname, specifically designed for Microservices. \textit{Mocking-Assisted Seed Execution}, \textit{Distributed Tracing}, \textit{Seed Refresh} and \textit{Pipeline Parallelism} approaches are adopted to address the environmental complexities and dynamics of Microservices and improve the efficiency of fuzzing.
\toolname has been successfully implemented and deployed in \company ~\footref{footnotetargetcompany}, a prominent FinTech company. Its performance has been evaluated in three distinct industrial scenarios: normalized fuzzing, iteration testing, and taint verification. 

Throughout five months of operation, \toolname has diligently analyzed a substantial codebase, consisting of 261 Apps with over 74.6 million lines of code (LOC). The framework's effectiveness is evident in its detection of 5,718 potential quality or security risks, with 1,764 of them confirmed and fixed as actual security threats by software specialists. Moreover, \toolname significantly increased line coverage by 12.24\% and detected new paths by 38.42\% in the iteration testing.
\end{abstract}

\copyrightyear{2024}
\acmYear{2024}
\setcopyright{acmlicensed}\acmConference[ICSE-SEIP '24]{46th International Conference on Software Engineering: Software Engineering in Practice}{April 14--20, 2024}{Lisbon, Portugal}
\acmBooktitle{46th International Conference on Software Engineering: Software Engineering in Practice (ICSE-SEIP '24), April 14--20, 2024, Lisbon, Portugal}
\acmDOI{10.1145/3639477.3639723}
\acmISBN{979-8-4007-0501-4/24/04}

\begin{document}
\maketitle
\section{Introduction}

Fuzzing is a widely adopted technique in the software industry to enhance security and ensure software quality. Several fuzzers such as AFL~\cite{afl}, libFuzzer~\cite{libFuzzer}, honggfuzz~\cite{honggfuzzer} and their extensions~\cite{directgreyfuzz,markov,collAFL,steer,NEZHA,10.1145/3551349.3556929} have successfully uncovered numerous bugs in both open-source and commercial programs~\cite{googleData}. 
However, the existing fuzzers primarily focus on \textit{monolithic software} wherein the program's components or functions are tightly coupled.
They are not readily applicable to fuzzing Microservices software, which has emerged as one of the most popular software architectures in the industry~\cite{newman2021building, microserviceswiki}. The adaptation of fuzzing techniques for microservices faces obstacles due to the differences between monolithic and distributed software architectures~\cite{zhang2022fuzzing,10.1145/3585009,10298503,uFuzz}.

\textbf{Inconsistency.} Existing fuzzers struggle to adapt to the intricate Microservices in the industry, primarily due to their inability to address Microservices' \textit{inconsistency} issue that refers to the execution paths that are not repeated or predictable. The inconsistency arises from the unpredictable runtime logic of Microservice frameworks~\cite{yannis}. Additionally, the independent development and redeployment of Microservice Applications (abbr. Apps) can result in temporary failures in cross-App invocations~\cite{zhong2023}. These factors contribute to the inconsistency observed in Microservices, posing a challenge for enabling existing fuzzers in Microservices.

\textbf{Communication.} Given that Microservice Apps are often deployed separately in different containers, the communication overhead between the target Apps and the fuzzer is essential for efficiency~\cite{newman2021building, microserviceswiki}. Traditional fuzzers often overlook the network consumption considerations specific to Microservices. 
Coordinating different fuzzing components and hiding the substantial time delay caused by such architectural disparities remains a daunting and challenge task for Microservice fuzzers.

\textbf{Applicability.} Existing fuzzers fail to meet the cost requirements associated with the extensive code bases prevalent in the industry. For instance, the Microservices software in \company~\footnote{\company is a prominent FinTech company that serves billions of users worldwide and processes 1 million user requests per minute on average.\label{footnotetargetcompany}} comprises more than 3,000 Microservices Apps and encompasses hundreds of millions of lines of code. 
To minimize the cost of fuzzing, it is crucial to determine when to terminate the fuzzing process across thousands of Apps and assess the impact of termination on fuzzing effectiveness.

In this paper, we introduce \toolname, a novel fuzzing framework specifically designed and implemented to tackle the challenges associated with Microservices architecture. 
For \textbf{Inconsistency}, we overcome the oracles of environmental complexities and dynamics using \textit{Mocking-Assisted Seed Execution} and \textit{Seed Refresh} techniques, and propose a Microservice testing harness by virtue of the above two and \textit{Distributed Tracing} techniques. To address \textbf{Communication}, we decouple different fuzzing components, making each of them parallel work in \textit{pipelines}, to mitigate the low efficiency caused by the across-App network communication in Microservice software. Besides, for \textbf{Applicability}, we utilized the ecological idea of the study~\cite{progress} to fuzz the industrial-level Microservice software, greatly saving the CPU consumption, and make our Microservice fuzzing framework cut out for large-scale industrial scenarios: we take \textit{Iteration Testing} and \textit{Taint Verification}, two important software quality assurance processes used in enterprises, as examples to demonstrate \toolname's applicability in industrial fields.

We successfully deployed \toolname in \company to evaluate a total of 261 representative Apps from a pool of over 3,000 Microservice Apps. These Apps spanned various business domains, including e-commerce, insurance, and investments, and encompassed commonly used Apps such as \textit{account}, \textit{trade}, and \textit{payment}. These Apps generated significant traffic and had substantial representations within the FinTech Microservices. Over five months, \toolname uncovered 5,718 potential security risks, out of which 1,764 risks were confirmed through internal cybersecurity exercises, highlighting the effectiveness of the tool in identifying genuine vulnerabilities. Furthermore, \toolname significantly increased line coverage by an average of 12.24\% across the tested applications.  \toolname also enhanced the richness of testing in the Iteration Testing scenario by 38.42\% more program paths.

In summary, we make the following contributions:
\begin{itemize}[leftmargin=0.3cm]

 \item We have developed an innovative fuzzing framework called \toolname specifically designed for Microservices Software. Our framework, \toolname, is tailored to effectively and efficiently fuzz Microservice software deployed in cloud environments.

\item We have employed the \textit{Mocking-Assisted Seed Execution}, \textit{Distributed Tracing} and \textit{Seed Refresh} techniques to tackle the complexities and dynamics of Microservice software and facilitate Microservice fuzzing. Furthermore,  we have leveraged \textit{Pipeline Parallelism} to decouple fuzzer components and improve the overall efficiency of Microservice fuzzing.

\item \toolname has been successfully deployed at \company and utilized for evaluating hundreds of Microservices Apps. Over five months, it has effectively identified and confirmed 1,764 quality issues and security threats, resulting in significant improvements to program coverage with an average increase of 12.24\%. Moreover, \toolname provides support for various program assurance processes, as it exhibits 38.42\% more program paths in the iteration testing scenario.

\end{itemize}

\section{Background}

In this section, we describe some key features of Microservices and some challenges posed by these features to fuzzing.

\subsection{Microservices}\label{subsec_microservice}

\begin{figure}[t]
  \centering
  \includegraphics[width=.4\textwidth]{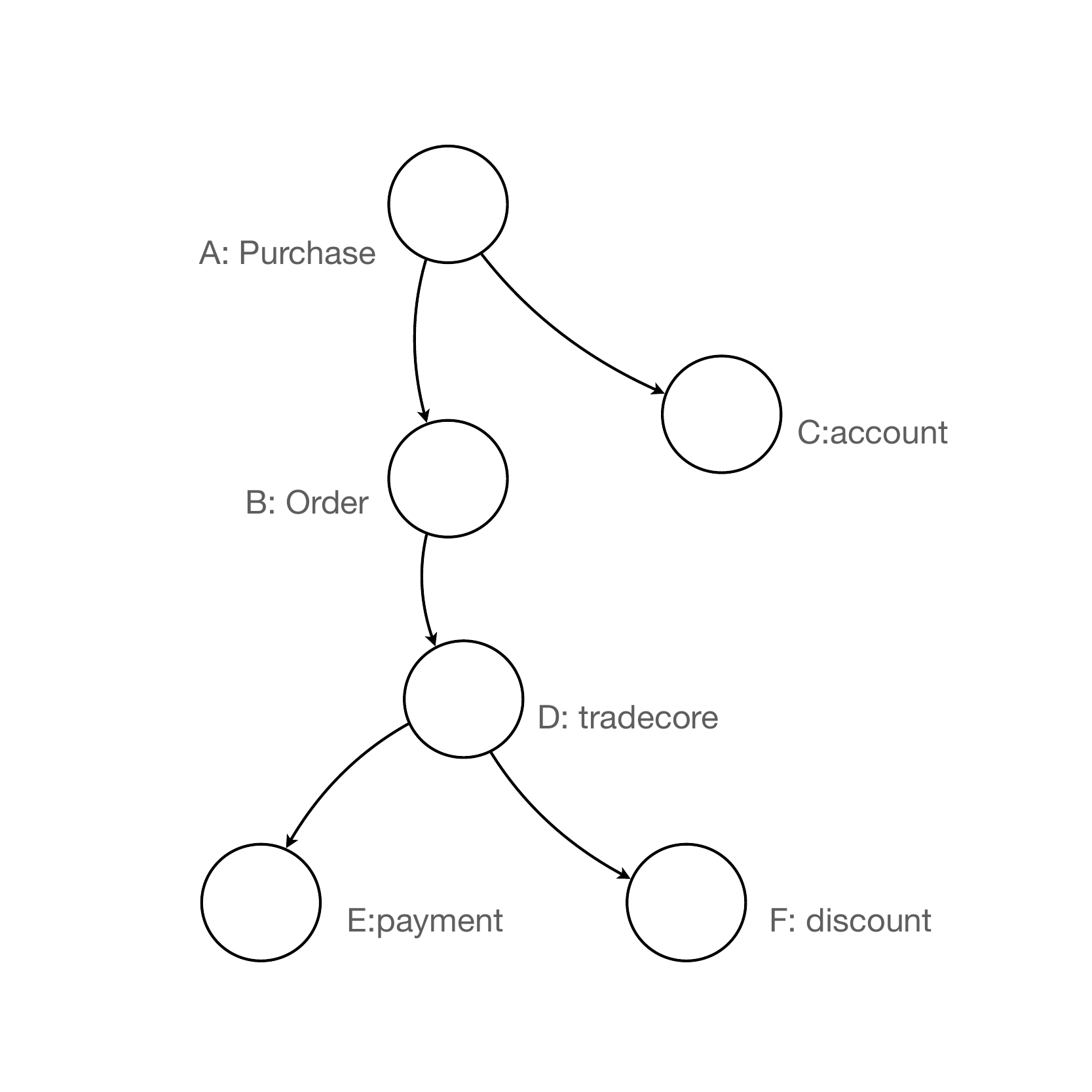}
  \caption{An example of Microservices and its Apps.}
  \label{fig:microservices_demo}
\end{figure}

Microservices~\cite{newman2021building, microserviceswiki} is an architecture pattern that arranges an application as a collection of loosely coupled, fine-grained services, communicating through network protocols. 
In this paper, a system implemented with this architecture is called \textit{Microservices Software}, while each service 
 working as a component is called \textit{Microservices Applications} (abbr. App). Figure~\ref{fig:microservices_demo} depicts an example of Microservices Software for processing goods purchases. The software consists of six Microservices Apps, each with a specific and bounded functionality. For instance, service \textit{C} handles account-related transactions, while service \textit{B} manages order-related transactions. This architecture promotes modularity and specialization within the system.

Microservices has several key features:
\begin{itemize}[leftmargin=0.3cm]
    \item \textbf{Independence}. Each App usually is developed and maintained by an independent team and is independently deployed. 
    This makes the development of a Microservice software more flexible than a monolithic software.

    \item \textbf{Interactivity}. While Microservices Apps can be developed and deployed independently, they still require communication with each other through network protocols like HTTP and RPC (Remote Procedure Call) to collaborate on user requests. In other words, from the perspective of effectively handling user requests, there is interdependence and interactivity among the Microservices Apps.

    \item \textbf{Relative Complexity}.
    In contrast to a monolithic software with equivalent functionality, each individual Microservices App has a smaller size and lower complexity. However, when considering the Microservices Software as a whole, it is different. Microservices in \company typically comprise thousands of Apps, encompassing hundreds of millions of lines of code. This indicates that the overall complexity is considerably high.

    \item \textbf{Rapid Evolution}.
    Development-independence of each App makes Microservice Software (as a whole) always be in a stage of evolution in industrial scenes, even if each App is relatively stable. In \company, for instance, there are thousands of merge requests (MR) being committed daily, and an average of one new version of an App is released every 4.3 days.
    
\end{itemize}

\subsection{Challenges}\label{subsec:challenges}
The above features of Microservices bring the convenience of deployment and development, but pose some challenges to fuzzing:

\begin{itemize}[leftmargin=*]
 \item \textbf{Challenge 1}. \textbf{How to enable a fuzzer for Microservices?} 
 
\textbf{Problem 1 (Inconsistency)} remains a significant obstacle for existing fuzzing techniques when it comes to testing complex Microservice software.
The fuzzing technique was originally designed for testing \textit{deterministic} programs, where multiple runs of an input consistently follow the same program path. For programs with \textit{uncertain} attributes (e.g., decision conditions determined by random numbers), traditional fuzzers are not suitable because the 
\textit{inconsistent}
program behaviors are hugely disruptive to the functions of \textit{seed selection} and \textit{trimming}, two important phases of the fuzzing workflow. To solve this issue, we adopt the \textit{Mocking-assisted seed execution} approach to approximately guarantee a consistent behavior (more detailed in \S\ref{mock}).\newline
\textbf{Problem 2 (Incomplete Program Coverage)} Fuzzing technique requires a total program coverage as the feedback of each fuzzing iteration. In monolithic software, coverage information is typically collected by directly reading the local memory. However, due to the distributed nature of Microservices, the coverage collection can only achieve multiple partial and disconnected coverages instead of obtaining a complete coverage view across all Microservices. (more detailed in \S\ref{trace}). In a word, the lack of microservice testing harness disables traditional fuzzing techniques to test Microservice software.

\item{\textbf{Challenge 2.}} 
\textbf{How to make Microservice fuzzing efficient?} Taking each App of an industrial-level Microservice software as a target, fuzzing faces two efficiency problems, caused by environmental complexities, dynamics, and architectural differences. \newline
\textbf{Problem 3 (Inconsistency-derived Inefficiency)} The adoption of the mocking-assisted technique helps address Problem 1, but it is not a perfect solution. Striving for strict consistency in fuzzing requires frequent replay of existing seeds, which can impact performance.
To balance consistency and performance in an efficient manner, we propose \textit{Seed Refresh \& Life-cycle Management} that involves periodically refreshing seed inputs during the fuzzing process (more detailed in \S\ref{refresh}).\newline
\textbf{Problem 4 (High Network Cost)} Due to inherent architectural differences, fuzzing an App deployed in a cloud environment poses distinct challenges compared to fuzzing desktop software or client-server-style protocol applications, e.g. AFL~\cite{afl}.
The fuzzer and the target App need to be deployed in separate containers, often residing on different hosts. This implies that communication between the fuzzer and the target App occurs through network protocols rather than the low-cost Inter-Process Communication (IPC) mechanism used by AFL and similar tools.

An App typically exposes interfaces
used for transmitting requests to other Apps and receiving their corresponding responses. 
To ensure efficiency, it is imperative to design a novel fuzzing framework that addresses the inefficiencies arising from network communication between Apps. To solve it, we redesign the fuzzing workflow by decoupling various fuzzing phases (more detailed in \S\ref{decouple}).  

\item{\textbf{Challenge 3.}} \textbf{How to scale Microservice fuzzing to real industrial scenarios?} Fuzzing is costly due to the continuous seed generation and execution. When to terminate it and whether the termination impacts the fuzzing effectiveness or not are important applicable metrics in the enterprise. For applicability, our approach is inspired by the concept of ecology-based fuzzing~\cite{progress} and implemented for the normalized fuzzing (more detailed in \S\ref{a1}). Besides, we explore the combination of our \toolname with other techniques to assure program quality and make two real practices in \company (more detailed in \S\ref{app}).

\end{itemize}

\section{Approaches and Designs}\label{aa}

We propose and develop a fuzzing framework targeting at Microservices Software, named \toolname. In this section, we will describe the overall architecture of \toolname and detail our solutions to address the identified challenges.

\subsection{\toolname's Architecture}\label{a0}

\begin{figure*}[!t]
\centering
\includegraphics[width=0.85\linewidth]{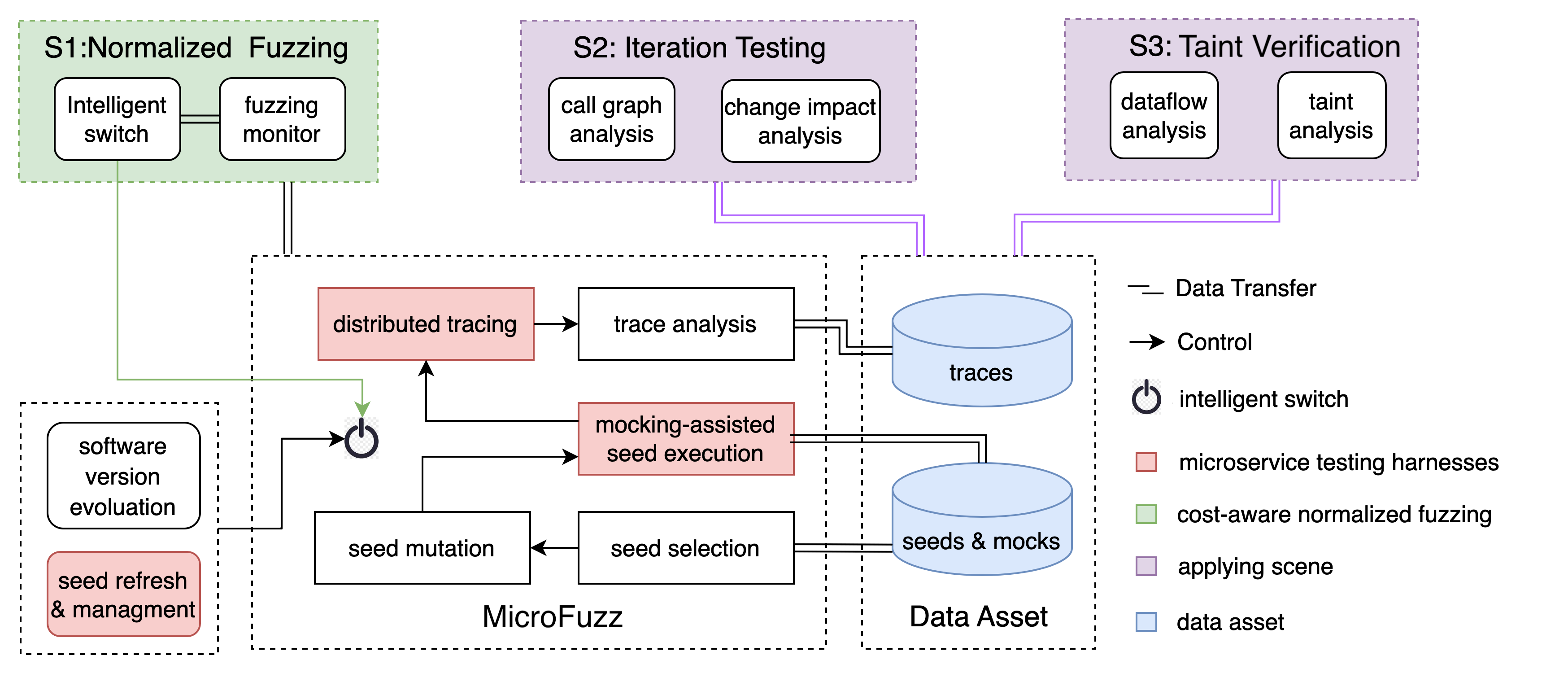}
\caption{\toolname's architecture and application scenarios.
}
\label{arch}
\end{figure*}

Figure~\ref{arch} shows the overall architecture of \toolname. It comprises an \textit{intelligent switch} and five core modules: \textit{seed selection}, \textit{seed mutation}, \textit{trace analysis}, \textit{mocking-assisted seed execution}, and \textit{distributed tracing}. While the first three modules are traditional components, the latter two are newly designed modules serving as microservice testing harnesses, specifically enabling microservice fuzzing.

\textcircled{1} 
\textit{Seed selection} module is responsible for selecting the optimal initial seeds from a large set. These seeds are obtained from real Internet traffic or generated by our fuzzing tool in previous iterations. 

\textcircled{2} \textit{Seed mutation} module is responsible for generating new test cases using mutation strategies commonly employed in traditional fuzzing techniques. 

\begin{itemize}[leftmargin=0.5cm]
    \item \texttt{Bit/Byte Mutation}: bit or byte-level flip 
    \item \texttt{Arithmetic Mutation}: add/subtract one on original integer
    \item \texttt{Interesting Replacement}: replace some bytes with the \textit{interesting} integer value belonging to (0, INT\_MAX)
    \item \texttt{Havoc Mutation}: set random bytes to random values; delete or duplicate some byte sequences
    \item \texttt{Splice Mutation}: splice two seeds at an arbitrary midpoint
\end{itemize} 

\textcircled{3} \textit{Mocking-assisted seed execution} module is responsible for executing a seed by invoking either real or mocked services (more detailed in \S\ref{mock}).

\textcircled{4} \textit{Distributed tracing} module is responsible for collecting comprehensive coverage information during the execution of a seed in a cluster environment (more detailed in \S\ref{dtrace}).

\textcircled{5} \textit{Trace analysis} module is responsible for analyzing the execution trace of a seed and deciding whether to store it in databases or not. Similar to traditional fuzzing techniques, only seeds that trigger new crashes or cover unique program locations are typically stored for further investigation and utilization.

\subsection{Mocking-Assisted Seed Execution}\label{mock}

Microservice software is highly complex and undergoes rapid evaluation due to frequent releases and interactivity-dependency among its component Apps. This inconsistency poses a challenge for traditional fuzzing techniques, making them unsuitable for testing Microservice software. To tackle this challenge, we utilize the mocking method~\cite{rockcode} to ensure approximate \textit{seed consistency}, which allows multiple runs of a seed to follow the same program path within a short time period. The consistency is maintained by replacing the values of each mocking point with the recorded values from the previous version. The process begins with static analysis, where three types of mocking points are identified and collected:

\begin{itemize}[leftmargin=0.5cm]
  \item \textbf{System Dependencies}. An App usually calls some library methods closely related to the operating system. Outputs of these methods are actually decided by the underlying OS, such as \texttt{Math.random()}  method and \texttt{System.currentTimeMillis()} method. We call them \textit{system dependencies}.

  \item \textbf{Internal Dependencies}. In an App, some static fields and fields of some singleton beans are usually used to maintain some global (internal) states. 
  For instance, the \texttt{counter} variable in Figure~\ref{internal} will be increased by one every time when \texttt{getValue()} method is called, making any seed that reaches this location be \textit{inconsistent} due to the different \texttt{counter} of each run.
  
  \item \textbf{External Dependencies}. One App usually refers to the exposed interfaces (e.g. RPC Reference of Spring) of other Apps to transfer some sub-requests to them and cooperate to finish user requests. Because the dependent Apps may release new versions during fuzzing, improper processing on these external dependencies also easily causes \textit{inconsistency} issues.
\end{itemize}
\label{26}

Corresponding to the above dependencies, we refer to the program locations involved as \textit{mocking points}, denoted as $\boldsymbol{PS}$, $\boldsymbol{PI}$, and $\boldsymbol{PE}$ for brevity. Using the static analysis approach presented in~\cite{rockcode}, we identify these mocking points and establish a consistent execution of seeds by following the steps outlined below.

\textbf{Mocking Points Collection.}
Given a Microservices App, all its mocking points $\bm{M}=\{m_1,m_2,\cdots\}$ are recursively gathered from its bytecode, using~\cite{rockcode}'s method: (1) $\boldsymbol{PS}$ are collected if the called functions are from or tainted from the third libraries;
(2) $\boldsymbol{PI}$ are collected if they are \textit{static/Spring bean} fields or tainted by any \textit{static/Spring bean} fields;
(3) $\boldsymbol{PE}$ are collected if they are the \textit{database access objects (DAO)} invoking the Mybatis mapper interfaces, or the \textit{RPC references} configured in the XML files via the tags of \texttt{sofa:reference} or \texttt{sofa:binding.tr.RPC} or tainted by DAOs and RPC references.

\textbf{Seed Record.}
\texttt{JVM} injector is utilized to record a seed's execution. For any seed $s$, the  input and output at each mocking point $m_i$ are recorded in format of $m_i^{s}=(input_i^{s}, output_i^{s})$, 
and its execution can be described using $\bm{M}^{s}=\{m_1^{s}, m_2^{s}\cdots\}$. Note that if a mocking point $m_k$ is not covered in that run, both its input and output will be set to \texttt{NULL} as $m_k^{s}$ = $(\texttt{NULL}, \texttt{NULL})$.

\textbf{Seed Replay.}
Based on the latest $\bm{M}^{s}=\{m_1^{s}, m_2^{s}\cdots\}$, a \texttt{JVM} injector is used to perform the value replacement in seed $s$'s replay: At each mocking point $m_i$, the \texttt{JVM} injector captures the current input value $input_{c_i}^{s}$ in real-time, and then $output_i^{s}$ will be directly returned, rather than be really executed, if and only if a same input is observed in this execution (i.e. $input_{c_i}^{s}=input_i^{s}$); Otherwise, a real execution will be performed, without any value replaced. 
We have established rules for assessing equality. If the features extracted from two inputs are categorized as similar (not necessarily identical), then we define them as equal.

\begin{figure}
\centering
\begin{minipage}[ht]{1\linewidth}
\centering
\lstset{frame=lrtb,
  language=Java,
  aboveskip=3mm,
  belowskip=3mm,
  xleftmargin=2mm,
  xrightmargin=2mm,
  showstringspaces=false,
  columns=flexible,
  basicstyle={\small},
  numberstyle=\tiny\color{gray},
  keywordstyle=\color{blue},
  commentstyle=\color{dkgreen},
  stringstyle=\color{mauve},
  breaklines=true,
  breakatwhitespace=true,
  tabsize=2
}
\begin{lstlisting}[keywordstyle=\color{blue!70},commentstyle=\color{red!50!green!50!blue!50},frame=shadowbox, rulesepcolor=\color{red!20!green!20!blue!20}]
public class InternalExample {
    private static Integer counter = 0;
    private static Map<String, String> kvmap = new HashMap();
    public static Integer getCounter() {
        return counter;
    }
    public static String getValue(String key) {
        counter++;
        return kvmap.getOrDefault(key, "");
    }
    public static void putValue(String key, String value) {
        kvmap.put(key, value);
    }
} 
\end{lstlisting}
\end{minipage}
\caption{An example of static fields influencing the internal state.}
\label{internal}
\end{figure}

\subsection{Distributed Tracing}\label{trace}

Problem 2 highlights the difficulty of achieving comprehensive coverage information in Microservice software, which is distributedly deployed in a cluster. Traditional fuzzing techniques struggle to accomplish their testing goals under such circumstances. To overcome this challenge, we propose the utilization of the \textit{distributed tracing} technique.

\begin{figure}[!t]
\centering
\includegraphics[width=0.8\linewidth]{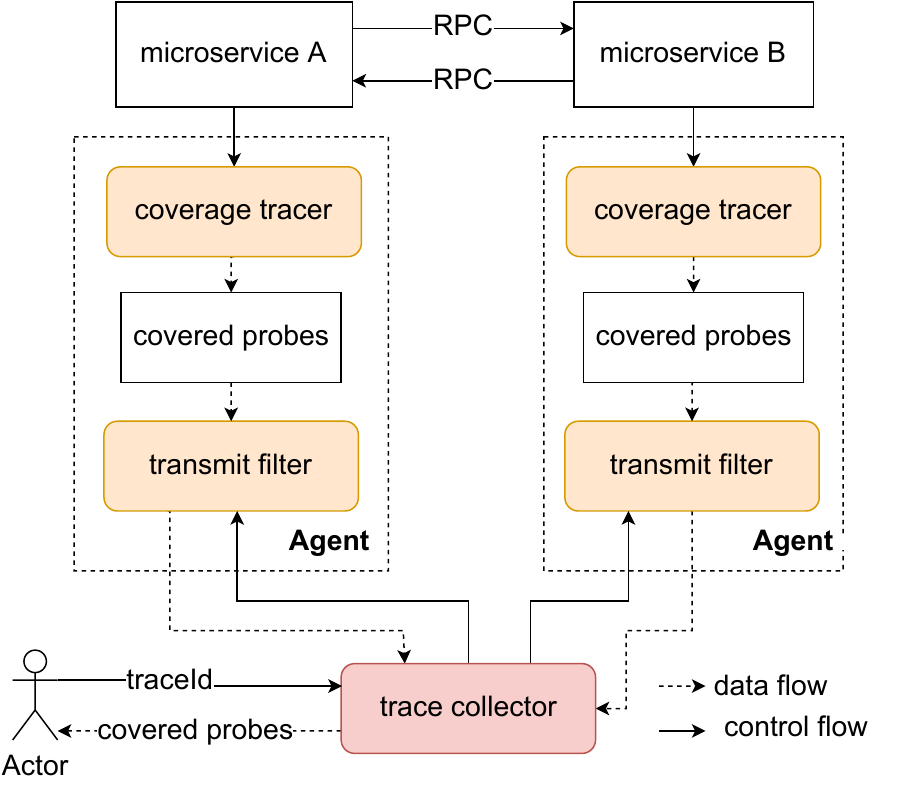}
\caption{Distributed tracing architecture.}
\label{dtrace}
\end{figure}

Our approach involves deploying agents throughout the cluster to collect coverage information during seed execution in Microservice software. 
Agents record covered probes and assign unique \textit{traceId} to request contexts. The recorded coverage is hashed into \textit{coverDigest} and stored locally. Unique \textit{coverDigest} is transferred to the central \textit{Trace Collector}, signaling the discovery of new program behavior (i.e., program path). Users can query the complete program coverage using \textit{traceId}. The \textit{Trace Collector} retrieves relevant covered probes from agents, splices them based on timing sequences, and generates an entire coverage description following the OpenTracing specification~\cite{opentrace}. Figure~\ref{dtrace} illustrates this process.

\subsection{Seed Refresh \& Life-cycle Management} 
\label{refresh}

In an industrial scene, Microservices exhibit high complexity and dynamics, where a single request to an App can trigger a sequence of service invocations. When focusing on a specific App, modifications to its upstream services can potentially impact its behavior. This is because the App relies on the output of upstream Apps as input to execute its own business logic. This situation is illustrated as Problem 1. In an enterprise setting, it is common to see hundreds to thousands of merge requests (MRs) being committed on a daily basis, further contributing to the dynamic nature of Microservices.
Consider a target App with $m$ upstream services, where each service undergoes an average of $n$ changes per day. As a result, the program behaviors of the target App can potentially change up to $mn$ times daily. To accurately capture and describe the program behaviors of the latest version of each App, it becomes necessary to refresh all the App's seeds in the seed database whenever the target App or its upstream Apps release new versions. This process may involve performing seed replay up to a maximum of $mn$ times per day.

Considering the frequent software version evolution and huge amount of seeds, $mn$-times \textit{seed re-execution} is definitely unacceptable in practice, as Problem 3 shows. Instead of this accurate but expensive method, we propose an industrially degraded approach. 
We make use of an \textit{event listener} and two \textit{crontabs} to approximately enable the \textit{Seed Refresh \& Life-cycle Management}. As Figure~\ref{arch}, triggered by \textit{Seed Refresh} or \textit{Software Version Evolution}, any seed in the seed database will be replayed by \textit{Seed Execution} module every 12 hours or once the given App releases a new version. 
During such replay, those seeds will be run on the latest version of the target App, and invocations to other Apps will be concretely transferred to these dependent Apps to get real responses instead of using the mocking value stored in the database (i.e. \textit{Mocking-Assisted Seed Execution}).
The real responses will then be used to update the mocking database.
Besides, \textit{Seed Life-cycle Management} module cleanups all the seeds, created three days ago, together with their outdated mocking values.

\subsection{Pipeline Parallelism} \label{decouple}

\begin{figure}[!t]
\centering
\includegraphics[width=\linewidth]{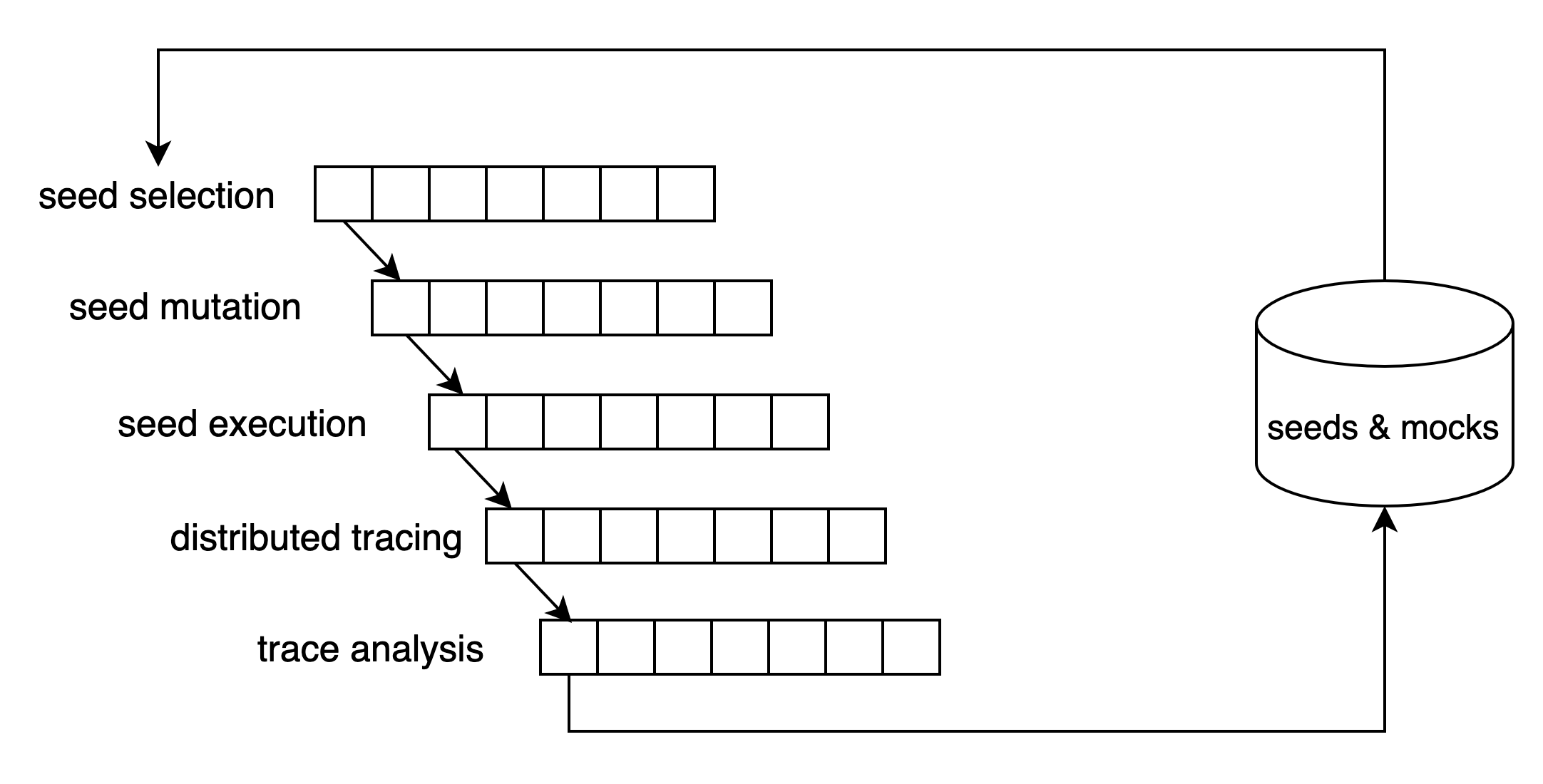}
\caption{Pipeline parallelism.}
\label{decouplexx}
\end{figure}

Fuzz testing typically involves several steps, such as seed mutation, seed execution, and coverage collection. In traditional monolithic fuzzing techniques, these steps are performed sequentially. This means that the next fuzzing iteration can only begin once the current iteration is complete and coverage has been obtained. However, Problem 4 highlights the presence of high network costs between Microservice Apps. The sequential nature of traditional fuzzing significantly hampers the fuzzing process and reduces its efficiency.
To address this issue, we have decoupled all the fuzzing phases, as depicted in Figure~\ref{decouplexx}. In this decoupled work mode, the fuzzing process no longer needs to wait for the return of distributed tracing and analysis results. This greatly improves the efficiency of the fuzzing process, as it can continue without delay.

\section{Applications}\label{app}
We have applied \toolname in practical industrial scenarios, including normal fuzzing scenario to find bugs and improve coverage (\S\ref{a1}), iteration testing scenario to find more behaviors manifested in an iteration (\S\ref{a2}), and taint verification scenario to confirm data leakage risks found by dynamic taint analysis (\S\ref{a3}). 
In this section, we show the details of these three application scenarios.

\subsection{S1: Normal Fuzzing Scenario}\label{a1}

Fuzz testing often wastes computational power on ineffective exploration, where no new coverage or crashes are found. Based on the insights gained from AFL's fuzzing exploration~\cite{afl}, it has been observed that most program locations are covered in the early stages, with only a few new discoveries made later on. Therefore, an \textit{Intelligent Switch} is introduced into \toolname to control core modules (e.g. \textit{Seed Scheduler} and others) and make power-aware decisions. Once this switch identifies the fuzzing progress trapped and stuck in the above inefficient loops, it will terminate the working of each fuzzing module immediately. The extended \textit{Fuzzing Monitor} collects and analyzes exploration information, while the switch is activated during software updates, inconsistent behavior, or user commands. It can be turned off by user commands or the \textit{Fuzzing Monitor} to optimize computational resource usage.

B\"{o}hme et al.~\cite{progress} draw a strong similarity between fuzzing exploration and species discovery in \textit{ecology}. Both involve mutation strategies, spatial distribution, and temporal dynamics. Statistical methods used in species discovery can be applied to answer key questions in fuzzing campaigns, such as the presence of undetected bugs, maximum program coverage achievable, and the number of additional test cases required to reach desired coverage.

In the normalized fuzzing scenario, the \textit{Fuzzing Monitor} is specifically designed to collect the information, shown in Table~\ref{tab:trackfactor}, during the whole fuzzing exploration. 

\begin{table}[h]
\centering
\caption{Tracking factors and estimation.}
\label{tab:trackfactor}
\begin{tabular}{clp{5cm}}
\hline
\multicolumn{2}{l}{\textbf{Tracking factors}}\\
\hline
$n$ & the number of seeds replayed \\
$\hat{S}$ & the number of all statements in target App\\
$S(n)$ & the number of statements covered by $n$ seeds\\
$f_1$ & the number of \textit{singletons*} detected\\
$f_2$ & the number of \textit{doubletons*} detected\\
$Q_0$ & the number of undiscovered statements,$Q_0=\hat{S}-S(n)$ \\
$Q_1$ & the number of statements only executed by \textit{singleton}\\
$C$ & the number of stored seeds with different \textit{coverDigest} \\
\hline
\multicolumn{2}{l}{\footnotesize * \textit{singleton}/\textit{doubleton} mean the seed whose \textit{coverDigest} has occurred once/twice.}\\
\end{tabular}
\end{table}

We utilize the method proposed in~\cite{progress} to estimate our framework.
\begin{itemize}
  \item $f_1/n$ represents the ratio or proportion of new crashes or coverage discovered.
  \item $U(n) = C(n)/(C(n) + (n-1)f_1^{2}/(2nf_{2}))$ represents the \textit{upper bound} of potential program coverage or vulnerabilities that can be discovered by continuous fuzzing exploration.
  \item ${S}(n+m) =S(n) + Q_0[1-(1-{Q_1}/({nQ_0 +Q_1}))^{m}])$ is the expected program coverage or vulnerability count after executing $m$ more seeds. 
\end{itemize}

The \textit{Fuzzing Monitor} utilizes statistical estimations to make decisions regarding the \textit{Intelligent Switch} behavior. If no new crashes or coverage are expected, indicated by either a low value of $f_1/n$ or $U(n) - \hat{S}(n+m)$ compared to a threshold $t$, the monitor turns the switch off. Conversely, if inconsistencies are detected during seed replay or software version evolution, the monitor turns the switch back on. The intelligent on-off control optimizes computing resources without compromising fuzzing effectiveness. Figure~\ref{normal} demonstrates the successful balance achieved, where resources are efficiently allocated while maintaining the ability to discover crashes and coverage during fuzzing.

\subsection{S2: Iteration Testing Scenario}\label{a2}
Software evolves by new feature import or logic update in each iteration. To plan tests for an iteration, two aspects should be considered. One is \textit{regression test} which reruns the prior test cases to ensure the previous functions performed as expected after a change~\cite{regression01}. Another is to test the given software as thoroughly as possible to ensure that the new features are free from any residual vulnerabilities before release.
From the above aspects, this section will detail the extensions of the proposed \textit{microservice fuzzing} technology and introduce its application in \textit{iteration testing} scenario.

\begin{figure*}[!t]
\centering
\includegraphics[width=0.95\linewidth]{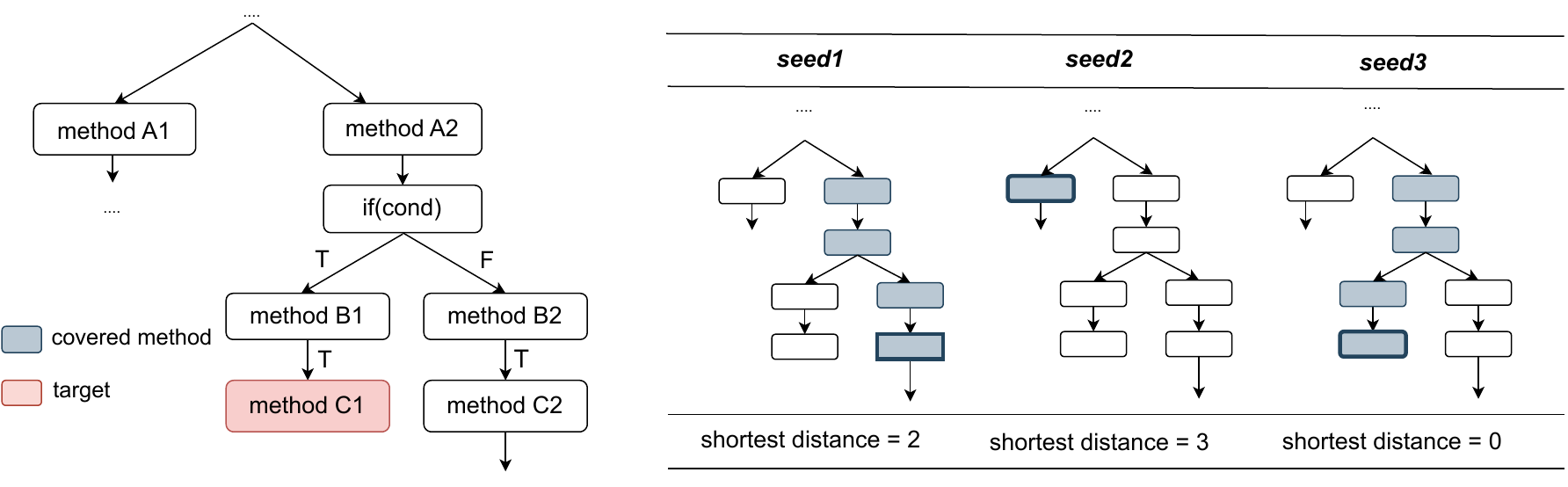}
\caption{Shortest distance calculation in directed fuzzing technique.}
\label{directx}
\end{figure*}

\textbf{Aspect of Regression Testing.}

Due to shorter software delivery life cycles and the presence of large test suites, re-running all test cases for a given software under test (SUT) after each change can be costly~\cite{regress_survey}. To address this, a technique called \textit{dynamic regression test selection (RTS)} is employed to select a subset of test cases from the test suite. These selected test cases are then executed on the latest version of the SUT to verify different program behaviors by comparing the new code changes with per-test execution traces~\cite{rst011,rts02,rts03}.
\toolname is extended with \textit{Change Impact Analysis}, as depicted in Figure~\ref{arch}, to associate each execution trace with line-, block-, or method-level coverage. These traces are indexed based on commit information, such as the \textit{commitId} or the name and branch of the Microservice App. This indexing allows for quick retrieval of relevant traces from a large set, enabling efficient test selection and execution. Further implementation details will be discussed in \S\ref{imple}.

For an iteration where a merge request (MR) is committed from branch \textit{A} to branch \textit{B}, \textit{regression testing} will be planned on test suite $T_A\cup T_B$, which is accomplished by following the steps:
\begin{itemize}
\item \textit{Change Impact Analysis} collects each different block $b$ between branch $A$ and branch $B$, and then puts it into $\bm{Diff}$. 
\item From \textit{traces} database, select each trace covering any different block $b\in \bm{Diff}$, and put it into set $R_A$ or $R_B$ if it was previously executed on branch \textit{A} or branch \textit{B}.
\item Based on the \textit{mapping} between \textit{traces} and \textit{seeds} database, the corresponding seed for each selected trace $t \in R_A \cup R_B$ is retrieved and stored in sets $T_A$ or $T_B$ accordingly, if the trace is derived from $R_A$ or $R_B$.
\end{itemize}

\textbf{Aspect of New Features Testing.}

In an iteration, testing efforts are focused on exploring new \textit{basic block} $n_i\in\bm{Diff}$ that have not been associated with any traces yet. \toolname utilizes directed fuzzing techniques and incorporates \textit{Call Graph Analysis} to obtain the call graph of the latest Microservice Software after each software version evolution. 
This is an incremental analysis process~\cite{zhao2023incremental}.
It also includes a \textit{Seed Scheduler} and \textit{Trace Analysis}, which select seeds based on the closest prior execution traces and calculate the shortest distances between methods using Dijkstra's algorithm~\cite{dijkstra}. These techniques optimize seed selection for effective testing and exploration of new features.

Figure~\ref{directx} illustrates the target method C1 for directed fuzzing exploration. The \textit{Seed Scheduler} module assigns decreasing priority to seed3, seed1, and seed2. Preference is given to seed3 and seed1 as they provide better coverage or proximity to the target method. This preference is also extended to the \textit{Trace Analysis} module during iteration testing, where seeds that can adequately cover or closely match the target are prioritized.

\subsection{S3: Taint Verification Scenario}\label{a3} 

Taint information plays a critical role in ensuring program quality and safeguarding data privacy within the industry. To capture taint information, dataflow and taint analysis techniques are commonly employed, allowing for the monitoring and tracing of data propagation from specific sources. However, existing traffic volumes often fall short in providing sufficient data for accurate verification. To overcome this limitation, our \toolname bolsters verification capabilities by significantly increasing the volume of traffic available for analysis.
In comparison to static taint analyzers with lower precision~\cite{taintr01,taintr02} and dynamic taint analyzers with higher costs~\cite{taintc,dytan02}, the combination of our \toolname with static analyzers has demonstrated its effectiveness and practicality. We have successfully validated this approach through two real industrial practices, and further results can be found in Table~\ref{result1}.

To establish the taint verification, \toolname works as follows:
\begin{itemize}
\item from \textit{seeds} database, \textit{Seed Selection} picks a seed $s_o$ whose prior execution trace reaches the \textit{source}'s method entry.
\item \textit{Seed Mutation} generates new seeds $\bm{s_k}=\{s_1, s_2, \cdots\}$ by only performing mutations on the \textit{source} parameter of $s_o$.
\item in each $\bm{s_k}$'s execution, \textit{Trace Analysis} observes the \textit{sink}'s value and denotes it as $sink_i$. Using the \textit{variable-controlling} approach, <\textit{source},\textit{sink}> is verified if $sink_i\neq sink_j$, $i\neq j$. Otherwise, the taint-relation is \textit{uncertain} (\textit{e.g.}, neither \textit{existence} nor \textit{inexistence} can be decided). 
\end{itemize}

\section{Implementation}
\label{imple}
\begin{table*}[!h]
\centering
\caption{Experimental results on 10 example Microservice Apps.}
\begin{tabular}[c]{*{11}{c}}
\hline
\toprule
\multirow{2}*{\textbf{Microservice}} & \multicolumn{5}{c}{\textbf{Effectiveness}}  & \multicolumn{4}{c}{\textbf{Efficiency}} \\
\cmidrule(lr){2-6}\cmidrule(lr){7-11}
~ & \textbf{\#LOC} & \textbf{COV (\%)} & \textbf{VUL}  & 
\textbf{New Traces} & \textbf{Taints}  &\textbf{\#TT (ms)} &\textbf{\#TS (ms)}  &\textbf{\#TO (ms)} &\textbf{\#TD (ms)} &\textbf{SAV (X)}  \\
\midrule
M1                       & 306K & +23.88\% & 3        & \#        & 9    & 55.26  & 3.06 & 172.32  & 117.32 &  0.46  \\
M2                       & 704K & +19.96\% & 2        & \#        & 90   & 66.64  & 2.74 & 139.64  & 73.62  &  0.89  \\
M3                       & 31K  & +8.54\%  & 0        & \#        & 28   & 54.35  & 3.18 &  93.26  & 39.18  &  1.38  \\
M4                       & 94K  & +17.83\% & 0        & \#        & 33   & 57.13  & 2.23 &  89.4   & 32.19  &  1.78  \\
M5                       & 818K & +21.05\% & 3        & \#        & 56   & 49.7   & 3.02 &  190.28 & 140.9  &  0.35  \\
M6                       & 156K & +17.48\% & 0        & 46.42\%   & 8    & 51.86  & 2.32 &  63.64  & 12.17  &  4.22  \\
M7                       & 273K & +18.29\% & 1        & 22.99\%   & 6    & 53.43  & 2.14 &  89.67  & 35.89  &  1.5   \\
M8                       & 504K & +13.03\% & 3        & 307.59\%  & 6    & 52.72  & 2.29 &  100.04 & 47.92  &  1.09  \\
M9                       & 75K  & +16.12\% & 0        & 176.72\%  & 467  & 49.52  & 2.79 &  97.38  & 47.14  &  1.07  \\
M10                      & 120K & +19.67\% & 0        & 109.09\%  & 15   & 62.2   & 2.22 &  71.2   & 9.02   &  6.89  \\
\hline
\textbf{10 Aggs}   & \textbf{24.96M} & \textbf{+17.59\%} & \textbf{12}  & \textbf{135.54\%}  & \textbf{718}    & \textbf{55.28} & \textbf{2.60} & \# & \# & \textbf{1.96} \\
\hline
\textbf{261 Aggs}   & \textbf{74.6M} & \textbf{+12.24\%} & \textbf{3831}  & \textbf{38.42\%}  & \textbf{5718}    & \textbf{52.72} & \textbf{2.89} & \# & \# & \textbf{2.67} \\
\bottomrule
\end{tabular}
\label{result1}
\end{table*}

We have successfully developed and deployed our proposed microservice fuzzing framework as a SOFABoot~\cite{sofa} microservice App, named \toolname. It comprises primarily of 263K lines of Java code.
\toolname can receive requests from either human users or CI/CD platforms, and it will commence fuzzing the designated Microservice App once receives a request.
In the following section, we will provide a thorough exposition of several pivotal components.

\textbf{Seeds \& Mocks Storage.} 
In \toolname, \textit{seeds} and \textit{mocks} databases serve as repositories for the respective seed and concomitant mock data. Similar to the fuzzing queue, the \textit{seeds} database facilitates frequent CRUD (Create, Read, Update, and Delete) operations throughout the fuzz exploration process, including the data access of \textit{Seed Mutation} in each fuzzing iteration and the bulk updating/deleting of data in regards to \textit{seed refresh} and \textit{seed life-cycle management}.

To ensure storage durability and efficiency, our \toolname utilizes the Cache-Aside pattern~\cite{aside} in conjunction with OceanBase~\cite{oceanbase} and Redis~\cite{redis}. Seeds and mocks are stored using OceanBase as the main store and Redis as the cache. 
Read-through and write-through strategies are implemented to load data from OceanBase into Redis on demand. 
Seeds are initially grouped by the application information, such as name, branch, or commitId, of the corresponding Microservice Software in OceanBase. In Redis, each seed is stored with its associated mocks to optimize \textit{Distributed Tracing}. The framework uses the <coverDigest, seed> pair to associate coverage information with each seed in Redis, enabling faster Trace Analysis by querying coverDigest for new program coverage.

\textbf{Seed Management and Software Version Evolution.} 

In \toolname, the functionalities of \textit{seed refresh} and \textit{seed life-cycle management} are facilitated by an \textit{event listener} and two \textit{crontabs}, respectively. The periodic tasks of \textit{seed replay} and \textit{cleanup} are carried out with the aid of AntScheduler~\cite{antscheduler}. Furthermore, the feature of \textit{software version evolution} is implemented through a \textit{callback handler instance} that is embedded in the CI/CD pipeline and will trigger upon the merging of a new commit.

\textbf{Intelligent Switch.}

In \toolname, the \textit{Intelligent Switch} serves as the primary controller for fuzzing exploration. It receives signals, such as \textit{on}/\textit{off} commands, from various sources including users, software version evolution, and the \textit{Fuzzing Monitor}. This functionality is implemented using Distributed Resource Management (DRM)~\cite{drm}, which enables dynamic modification of the entrance state of each core module while the Microservice Software is running. By setting the state of an entrance to \textit{off}, the corresponding fuzzing module becomes inaccessible unless it is reset to \textit{on} again. This capability allows the \textit{Intelligent Switch} to initiate or terminate a fuzzing exploration in real time based on the received signals.

\textbf{Trace-Block Association.}

To establish associations between traces and blocks, which is crucial for both \textit{regression testing} and \textit{directed fuzzing}, we employ Geabase~\cite{geabase}, a graph database, in \toolname.
When each software version evolves, a new call graph will be stored in Geabase and named using its git commit information. This allows for the identification and differentiation of different versions of the Microservice App. Each seed is associated with node $n$ of this graph once it reaches the corresponding basic block of $n$ during execution on the \textit{commitId} version.

\textbf{Data Flow Monitor.}

To implement the lightweight dynamic taint verification approach discussed in \S\ref{a3}, a \texttt{JVM} injector is utilized as part of the \textit{Distributed Tracing} process. This injector is responsible for gathering the values of service inputs and data access objects (DAOs) during the execution of the system. The purpose of this is to verify the taint relations between services or between services and the database. These taint relations are represented as <source, sink>, where the source indicates the taint position of the inputs (such as the parameters of a service), and the sink represents either the same position or a database operation. SOFAMQ~\cite{sofamq}, a distributed message middleware that builds upon RocketMQ~\cite{rocketmq}, is employed to decouple the key fuzzing modules, allowing the seamless and concurrent operations of the key fuzzing modules.

\section{Evaluation}

In this section, we present our evaluation setup and performance of our tool, \toolname, corresponding to the 3 key challenges defined in \S~\ref{subsec:challenges}.
That's to say, we try to answer three questions:
\begin{itemize}[leftmargin=0.3cm]
    \item Q1: Can \toolname effectively fuzz Microservice Apps deployed in an industrial environment, and what is its performance?

    \item Q2: How efficient \toolname is at fuzzing Microservice Apps?

    \item Q3: How does \toolname perform when applied in other scenarios except normal fuzzing?
\end{itemize}

It is worth noting that, to the best of our knowledge, \toolname is likely the first practical fuzzer for industrial-level Microservice software. As a result, when establishing baseline comparisons, we rely on user-provided data and feedback. Indeed, checking all the results from \toolname can be a challenging task, even for developers. This is primarily due to the large volume of data generated by the fuzzing process, often reaching thousands of entries for each Microservice App. Reviewing and analyzing such a vast amount of data can be time-consuming and resource-intensive.

\subsection{Setup}

We have deployed \toolname on a cluster of elastic cloud instances, each equipped with eight 2.5GHz cores and 32GB of RAM. Within the same cloud environment, we have deployed over 3,000 targeted Microservice Apps. \toolname has been running continuously for more than five months under this specific setup.

\subsection{Q1 - Effectiveness}

\begin{figure}[!t]
\centering
\includegraphics[width=\linewidth]{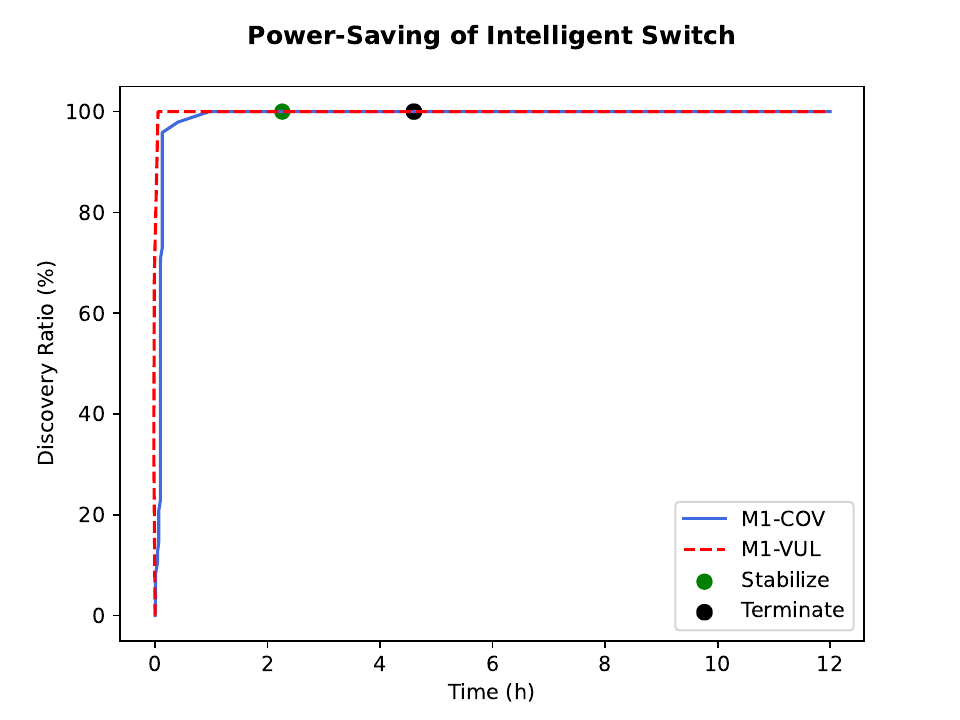}
\caption{Experimental results of Intelligent Switch.}
\label{normal}
\end{figure}

\textbf{Vulnerability Discovery Scenario.}

In order to assess the efficacy of \toolname in fuzzing Microservice Apps within an industrial cloud environment, we conducted a random selection process. Out of the over 3,000 Apps available, we chose 261 Apps to serve as our fuzzing targets. These selected Apps hold significant importance as they are fundamental, high-traffic, and heavily relied upon by other Apps in the ecosystem. Compared with others, they are more mature and fully tested in various business domains.

\begin{table}[t]
\centering
\caption{Vulnerability types on 261 target Microservice Apps.}
\label{tab:vul_stat}
\begin{tabular}{c|cc}
\hline
\textbf{Vul Type} & \textbf{Vul Detail} & \textbf{Count} \\
\hline
\multirow{2}*{Biz\_Vul} & Biz\_Exception & 254  \\
~ & Biz\_Error & 3548 \\
\cline{1-3}
\multirow{5} * {Sys\_Vul} & Null Pointer Exception  & 3 \\
~ & SQL Exception & 22\\
~ & NumberFormatException & 1\\
~ & UncleardThrowableException & 1\\
~ & IOException & 2\\
\hline
\end{tabular}
\end{table}

After five months of intensive fuzzing, 3,831 exception reports were uncovered.

Table~\ref{tab:vul_stat} presents a breakdown of the reported exception types, revealing Biz\_Error (a type of business error) as the most prevalent. 
Additionally, critical exceptions such as Null Pointer Exception have been identified during the evaluation process.
Most of Sys\_Vul exceptions have been patched by Apps' developers.

Furthermore, we specifically chose 10 sample Apps to present a more comprehensive overview of the effectiveness of \toolname, as shown in Table~\ref{result1}. 
These Apps have a line of code (LoC) ranging from 31,000 to 818,000.
The added code coverage achieved by \toolname ranges from 8.54\% to 23.88\%, with an average of 17.59\%, and 12 exceptions were found among these 10 Apps.
In total, the collective code coverage of the 261 Apps is 12.24\% on average, and 3,831 exception reports have been submitted.

Overall, we conclude that \toolname is effective in improving code coverage and discovering potential vulnerabilities.

\subsection{Q2 - Efficiency}

In addition to effectiveness, we also evaluated the efficiency of \toolname. For illustration purposes, we still take the selected 10 Apps as examples and present the results in Table~\ref{result1}.
The \textit{\#TT} column 
denotes the average time taken for  \textit{Distributed Tracing}, which is responsible for collecting runtime information during the fuzzing process for each App.
The average time duration for \textit{Distributed Tracing} across the 10 Apps is 52.72ms, ranging from 49.7ms to 66.64ms. This duration includes both the \textit{coverDigest} query and crash capture procedures.
\textit{\#TS} denotes the time
consumption of Distributed Tracing spent for data collection at each service entry or data access point. Each time takes around 2\textasciitilde3 ms on average, which proves our Distributed Tracing technique is efficient in the industrial scene.
The results demonstrate that our Distributed Tracing technique efficiently collects runtime information while maintaining the fuzzing performance at a satisfactory level.

The columns labeled \textit{\#TO} and \textit{\#TD} represent the total time required for a single fuzzing iteration with and without \textit{Pipeline Parallelism} respectively. The \textit{SAV} column denotes the improved efficiency achieved through \textit{Pipeline Parallelism}, calculated as SAV = \textit{\#TD}/\textit{\#TO} - 1. 
The efficiency improvement achieved through the use of Pipeline Parallelism varies. For example, App M6 and M10 experienced a 4.22x and 6.89x improvement. Conversely, the improvement for App M1 was relatively modest, with only a 0.46x increase. The difference comes from each App's characteristics. On average, however, the efficiency improvement across all Apps was notable, with a factor of 2.67x improvement achieved when employing the Pipeline Parallelism mechanism.

We evaluated the resource efficiency of the \textit{Intelligent Switch} mechanism, which optimizes computing resources by assessing the need for further fuzzing based on monitoring results. 
To demonstrate its effectiveness, we conducted a comparative study using M1 fuzzing as a case study, as depicted in Figure~\ref{normal}.
In this study, we performed two experiments: one with the Intelligent Switch enabled and another with it disabled. We plotted the changes in the proportion of discovered vulnerabilities over time (represented by the red line) and the code coverage ratio (represented by the blue line).
With the Intelligent Switch enabled, the fuzzer terminated after 4.6 hours as the mechanism determined that no new paths could be discovered, thus conserving computational resources. Conversely, in the control experiment without the Intelligent Switch, the fuzzer continued running, but no new paths were found, resulting in wasted resources.
On average, enabling the Intelligent Switch allowed for a 61.7\% reduction in computing resources while maintaining virtually no loss in path coverage. 
We have observed similar phenomena across nine other Apps as well.
This showcases the significant resource savings achieved by leveraging the effectiveness of the Intelligent Switch mechanism.

In summary, our adapted mechanism incorporating \textit{Distributed Tracing}, \textit{Pipeline Parallelism}, and \textit{Intelligent Switch} demonstrates high efficiency in terms of both time and computing resources. This outstanding performance establishes \toolname as an efficient tool for fuzzing Microservice Apps within an industrial cloud environment.

\subsection{Q3 - Applicability}
We assessed the applicability of \toolname in two additional scenarios: iteration testing and taint verification.

\begin{figure}[!t]
\centering
\includegraphics[width=\linewidth]{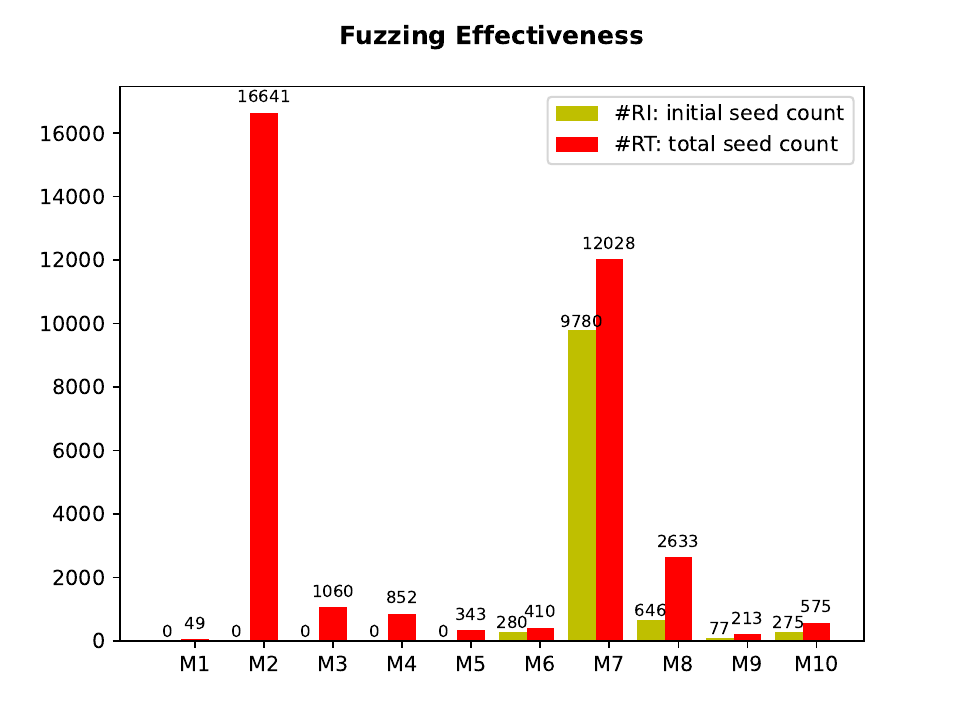}
\caption{Experimental results in iteration testing scenario.}
\label{direct}
\end{figure}

\textbf{Iteration Testing Scenario.}

When a code change is made to a Microservice App, it is crucial to generate relevant test cases that cover the modified code blocks to uncover any potential bugs, called \textit{iteration testing}. To address this, \toolname leverages directed fuzzing techniques.
In this specific use case, we evaluate \toolname's ability to generate new traces specifically targeting the modified code segments. To achieve this, we select traffic data that includes the code blocks affected by the changes as initial fuzzing seeds. We then analyze the number of newly discovered traces produced by \toolname as a measure of its effectiveness in capturing the behavior of the modified code segments.

Figure~\ref{direct} presents the average count of newly generated traces during the fuzzing process of the 10 selected Apps using \toolname, as depicted in Table~\ref{result1}.
\textit{\#RI} represents the count of initial seeds, while \textit{\#RT} indicates the number of traces discovered during the fuzzing process, including the initial seeds.
Taking the example of the M6 App, there were initially 280 seeds, and after the fuzzing process, a total of 410 traces were discovered. This indicates that \toolname uncovered 130 new paths, resulting in an effectiveness rate of 46.24\%.
The average effectiveness of the M6-M10 Apps, as depicted in the \textit{Traces} column of Table~\ref{result1}, was computed. Overall, \toolname demonstrated the capability to discover 38.42\% of new paths during each App's iteration.
It is important to note that the M1-M5 Apps do not have historical traffic data covering the changed code blocks. Therefore, in cases where a new functional module is introduced in a single iteration, the initial seed numbers for these modules in Figure~\ref{direct} would be 0, and there would be no corresponding effectiveness measure data in Table~\ref{result1}.

In summary, \toolname demonstrates its effectiveness in iteration testing by successfully identifying new coverage paths.

\textbf{Taint Verification Scenario.}

In this particular scenario, we utilized \toolname to validate the findings of static taint analysis, a technique prone to false positives. The testing results are presented in Table~\ref{result1}, where the \textit{Taints} column represents the cumulative number of taint-relations confirmed by \toolname. Additionally, the \textit{TS} column indicates the time required for \textit{Distributed Tracing} to collect data at each service entry or data access point.

Overall, \toolname identified a total of 5,718 unique possible instances of quality issues and security risks, of which 1,764 were confirmed by software specialists. These results demonstrate the effectiveness of \toolname in taint verification scenarios, enhancing the overall accuracy of the analysis process.

\section{Related Work}

\textbf{Fuzzers for Monolithic Software.} Representative fuzzers such as AFL~\cite{afl}, libFuzzer~\cite{libFuzzer}, and honggfuzz~\cite{honggfuzzer} provide guidance to other fuzzers. 
The typical fuzzing workflow consists of four phases: seed scheduling, seed mutation, seed execution, and seed selection. Various designs have been incorporated into these phases. 
B\"{o}hme et al.~\cite{markov} treat fuzzing exploration as a Markov process and propose a novel seed scheduling strategy that prioritizes seeds exploring low-frequency paths for further mutations. 
\textsc{FairFuzz}~\cite{fairfuzz} introduces new seed mutation operations, such as overwriting, deleting, and inserting, to enhance testing. 
\textsc{AFLGo}~\cite{directgreyfuzz} devises a seed selection strategy that favors seeds with execution paths closer to the target locations in each fuzzing iteration. Similar ideas are also explored in~\cite{steer}. 
Gan et al.~\cite{collAFL} address path collision issues in AFL by correcting path coverage calculations. 
Peng et al.~\cite{peng20191dvul} combined directed fuzzing with symbolic execution to reproduce 1-day vulnerabilities.
In recent years, many approaches have incorporated AI techniques into the seed mutation phase. For example, Zong et al.~\cite{fuzzguard} improve directed fuzzing efficiency by filtering out inputs predicted to be unreachable to targets. In~\cite{lstm}, an LSTM model is used to learn the mutable positions of inputs. Godefroid et al.~\cite{rnn2} employ an RNN to learn the grammar of program inputs using numerous test cases and use the learned grammar to generate new inputs. \textsc{NEUZZ}~\cite{neuzz} applies the concept of gradient descent to smooth the neural network model and significantly enhances program coverage by learning program branches.
With the recent enormous advances in Large Language Models (LLMs), TitanFuzz~\cite{deng2023titan} and FuzzGPT~\cite{deng2023fuzzgpt} have been proposed to directly leverage LLMs for fuzzing DL libraries.

Unlike other fuzzing frameworks, \toolname does not aim to improve existing fuzzing strategies. Instead, its main objective is to provide support and enable the effective implementation of these strategies specifically for Microservice software.

\textbf{Parallel Fuzzing.}
Existing approaches have made advancements in enhancing the performance of parallel fuzzing by focusing on improving the fuzzing strategy~\cite{Pham2021,UniFuzz,EnFuzz,PAFL,wang2021} or increasing the fuzzing speed~\cite{Xu2017,AFLplusplus-Woot20}. One common approach to enhance the fuzzing strategy is task partitioning. PAFL~\cite{PAFL}introduces an effective method to synchronize guiding information and statically divide fuzzing tasks based on branching information to minimize overlap between instances. AFLEdge~\cite{wang2021} utilizes static analysis to dynamically create exclusive and evenly weighted fuzzing tasks. Another strategy to improve fuzzing is through ensemble fuzzing~\cite{EnFuzz} or collaborative fuzzing~\cite{Cupid}, where the strengths of different fuzzers are combined. By fuzzing the same target with multiple fuzzers and sharing their progress, overall performance can be improved. EnFuzz~\cite{EnFuzz} designs heuristics to evaluate fuzzer diversity and selects the most diverse subset for ensemble fuzzing through efficient seed synchronization. Cupid~\cite{Cupid} proposes a collaborative fuzzing framework that automatically discovers the optimal combination of fuzzers for a target. One challenge in parallel fuzzing is the operating system bottleneck. Xu et al.~\cite{Xu2017} address this by introducing new operating primitives that enhance scalability and performance in parallel fuzzing, mitigating file system contention and scalability issues. 

A recent endeavor in parallelizing fuzzing involved the redesign of parallel fuzzing using a microservice architecture, as demonstrated in {{\textmu}FUZZ}~\cite{uFuzz}. It aimed to use the CPU power in the distributed cloud for fuzzing monolithic software by enhancing I/O operations and eliminating synchronization. 

Nevertheless, our \toolname is uniquely designed to address the specific challenges of identifying potential issues in Microservice software. As a result, the challenges we encounter differ from those encountered in other fuzzing frameworks.

\section{Conclusion}

This paper presents the first comprehensive study on Microservice fuzzing, which differs significantly from state-of-the-art fuzzing techniques for \textit{monolithic} software. To facilitate Microservice fuzzing, we propose an efficient fuzzing framework named \toolname incorporating innovative approaches such as \textit{Mocking-Assisted Seed Execution}, \textit{Distributed Tracing}, \textit{Seed Refresh}, and \textit{Pipeline Parallelism} to enhance fuzzing efficiency.
After running on thousands of Apps in \company, \toolname identified 5,718 potential quality or security risks, out of which 1,764 are confirmed. Furthermore, \toolname significantly enhances line coverage by 12.24\% and detects new paths by 38.42\% in the iteration testing.

\begin{acks}

This work is supported by Ant Group. 

\end{acks}

\bibliographystyle{ACM-Reference-Format}
\bibliography{sample-base}


\begin{thebibliography}{57}


\ifx \showCODEN    \undefined \def \showCODEN     #1{\unskip}     \fi
\ifx \showDOI      \undefined \def \showDOI       #1{#1}\fi
\ifx \showISBNx    \undefined \def \showISBNx     #1{\unskip}     \fi
\ifx \showISBNxiii \undefined \def \showISBNxiii  #1{\unskip}     \fi
\ifx \showISSN     \undefined \def \showISSN      #1{\unskip}     \fi
\ifx \showLCCN     \undefined \def \showLCCN      #1{\unskip}     \fi
\ifx \shownote     \undefined \def \shownote      #1{#1}          \fi
\ifx \showarticletitle \undefined \def \showarticletitle #1{#1}   \fi
\ifx \showURL      \undefined \def \showURL       {\relax}        \fi
\providecommand\bibfield[2]{#2}
\providecommand\bibinfo[2]{#2}
\providecommand\natexlab[1]{#1}
\providecommand\showeprint[2][]{arXiv:#2}

\bibitem[roc(2021)]%
        {rocketmq}
 \bibinfo{year}{2021}\natexlab{}.
\newblock \bibinfo{booktitle}{\emph{Apache RocketMQ}}.
\newblock
\urldef\tempurl%
\url{https://rocketmq.apache.org/}
\showURL{%
\tempurl}


\bibitem[Alibaba(2021)]%
        {oceanbase}
\bibfield{author}{\bibinfo{person}{Alibaba}.} \bibinfo{year}{2021}\natexlab{}.
\newblock \bibinfo{booktitle}{\emph{OceanBase}}.
\newblock
\urldef\tempurl%
\url{https://dbdb.io/db/oceanbase}
\showURL{%
\tempurl}


\bibitem[Antoniadis et~al\mbox{.}(2020)]%
        {yannis}
\bibfield{author}{\bibinfo{person}{Anastasios Antoniadis},
  \bibinfo{person}{Nikos Filippakis}, \bibinfo{person}{Paddy Krishnan},
  \bibinfo{person}{Raghavendra Ramesh}, \bibinfo{person}{Nicholas Allen}, {and}
  \bibinfo{person}{Yannis Smaragdakis}.} \bibinfo{year}{2020}\natexlab{}.
\newblock \showarticletitle{Static Analysis of Java Enterprise Applications:
  Frameworks and Caches, the Elephants in the Room}. In
  \bibinfo{booktitle}{\emph{Proceedings of the 41st ACM SIGPLAN Conference on
  Programming Language Design and Implementation}} (London, UK)
  \emph{(\bibinfo{series}{PLDI 2020})}. \bibinfo{publisher}{Association for
  Computing Machinery}, \bibinfo{address}{New York, NY, USA},
  \bibinfo{pages}{794–807}.
\newblock
\showISBNx{9781450376136}
\urldef\tempurl%
\url{https://doi.org/10.1145/3385412.3386026}
\showDOI{\tempurl}


\bibitem[Arzt et~al\mbox{.}(2014)]%
        {taintr01}
\bibfield{author}{\bibinfo{person}{Steven Arzt}, \bibinfo{person}{Siegfried
  Rasthofer}, \bibinfo{person}{Christian Fritz}, \bibinfo{person}{Eric Bodden},
  \bibinfo{person}{Alexandre Bartel}, \bibinfo{person}{Jacques Klein},
  \bibinfo{person}{Yves Le~Traon}, \bibinfo{person}{Damien Octeau}, {and}
  \bibinfo{person}{Patrick McDaniel}.} \bibinfo{year}{2014}\natexlab{}.
\newblock \showarticletitle{FlowDroid: Precise Context, Flow, Field,
  Object-Sensitive and Lifecycle-Aware Taint Analysis for Android Apps}. In
  \bibinfo{booktitle}{\emph{Proceedings of the 35th ACM SIGPLAN Conference on
  Programming Language Design and Implementation}} (Edinburgh, United Kingdom)
  \emph{(\bibinfo{series}{PLDI '14})}. \bibinfo{publisher}{Association for
  Computing Machinery}, \bibinfo{address}{New York, NY, USA},
  \bibinfo{pages}{259–269}.
\newblock
\showISBNx{9781450327848}
\urldef\tempurl%
\url{https://doi.org/10.1145/2594291.2594299}
\showDOI{\tempurl}


\bibitem[Bell and Kaiser(2014)]%
        {taintc}
\bibfield{author}{\bibinfo{person}{Jonathan Bell} {and} \bibinfo{person}{Gail
  Kaiser}.} \bibinfo{year}{2014}\natexlab{}.
\newblock \showarticletitle{Phosphor: Illuminating Dynamic Data Flow in
  Commodity Jvms}. In \bibinfo{booktitle}{\emph{Proceedings of the 2014 ACM
  International Conference on Object Oriented Programming Systems Languages \&
  Applications}} (Portland, Oregon, USA) \emph{(\bibinfo{series}{OOPSLA '14})}.
  \bibinfo{publisher}{Association for Computing Machinery},
  \bibinfo{address}{New York, NY, USA}, \bibinfo{pages}{83–101}.
\newblock
\showISBNx{9781450325851}
\urldef\tempurl%
\url{https://doi.org/10.1145/2660193.2660212}
\showDOI{\tempurl}


\bibitem[B\"{o}hme(2018)]%
        {progress}
\bibfield{author}{\bibinfo{person}{Marcel B\"{o}hme}.}
  \bibinfo{year}{2018}\natexlab{}.
\newblock \showarticletitle{STADS: Software Testing as Species Discovery}.
\newblock \bibinfo{journal}{\emph{ACM Trans. Softw. Eng. Methodol.}}
  \bibinfo{volume}{27}, \bibinfo{number}{2}, Article \bibinfo{articleno}{7}
  (\bibinfo{date}{jun} \bibinfo{year}{2018}), \bibinfo{numpages}{52}~pages.
\newblock
\showISSN{1049-331X}
\urldef\tempurl%
\url{https://doi.org/10.1145/3210309}
\showDOI{\tempurl}


\bibitem[B\"{o}hme et~al\mbox{.}(2017)]%
        {directgreyfuzz}
\bibfield{author}{\bibinfo{person}{Marcel B\"{o}hme},
  \bibinfo{person}{Van-Thuan Pham}, \bibinfo{person}{Manh-Dung Nguyen}, {and}
  \bibinfo{person}{Abhik Roychoudhury}.} \bibinfo{year}{2017}\natexlab{}.
\newblock \showarticletitle{Directed Greybox Fuzzing}. In
  \bibinfo{booktitle}{\emph{Proceedings of the 2017 ACM SIGSAC Conference on
  Computer and Communications Security}} (Dallas, Texas, USA)
  \emph{(\bibinfo{series}{CCS '17})}. \bibinfo{publisher}{Association for
  Computing Machinery}, \bibinfo{address}{New York, NY, USA},
  \bibinfo{pages}{2329–2344}.
\newblock
\showISBNx{9781450349468}
\urldef\tempurl%
\url{https://doi.org/10.1145/3133956.3134020}
\showDOI{\tempurl}


\bibitem[B\"{o}hme et~al\mbox{.}(2016)]%
        {markov}
\bibfield{author}{\bibinfo{person}{Marcel B\"{o}hme},
  \bibinfo{person}{Van-Thuan Pham}, {and} \bibinfo{person}{Abhik
  Roychoudhury}.} \bibinfo{year}{2016}\natexlab{}.
\newblock \showarticletitle{Coverage-Based Greybox Fuzzing as Markov Chain}. In
  \bibinfo{booktitle}{\emph{Proceedings of the 2016 ACM SIGSAC Conference on
  Computer and Communications Security}} (Vienna, Austria)
  \emph{(\bibinfo{series}{CCS '16})}. \bibinfo{publisher}{Association for
  Computing Machinery}, \bibinfo{address}{New York, NY, USA},
  \bibinfo{pages}{1032–1043}.
\newblock
\showISBNx{9781450341394}
\urldef\tempurl%
\url{https://doi.org/10.1145/2976749.2978428}
\showDOI{\tempurl}


\bibitem[Chen et~al\mbox{.}(2019)]%
        {EnFuzz}
\bibfield{author}{\bibinfo{person}{Yuanliang Chen}, \bibinfo{person}{Yu Jiang},
  \bibinfo{person}{Fuchen Ma}, \bibinfo{person}{Jie Liang},
  \bibinfo{person}{Mingzhe Wang}, \bibinfo{person}{Chijin Zhou},
  \bibinfo{person}{Xun Jiao}, {and} \bibinfo{person}{Zhuo Su}.}
  \bibinfo{year}{2019}\natexlab{}.
\newblock \showarticletitle{{EnFuzz}: Ensemble Fuzzing with Seed
  Synchronization among Diverse Fuzzers}. In \bibinfo{booktitle}{\emph{28th
  USENIX Security Symposium (USENIX Security 19)}}. \bibinfo{publisher}{USENIX
  Association}, \bibinfo{address}{Santa Clara, CA},
  \bibinfo{pages}{1967--1983}.
\newblock
\showISBNx{978-1-939133-06-9}
\urldef\tempurl%
\url{https://www.usenix.org/conference/usenixsecurity19/presentation/chen-yuanliang}
\showURL{%
\tempurl}


\bibitem[Chen et~al\mbox{.}(2023)]%
        {uFuzz}
\bibfield{author}{\bibinfo{person}{Yongheng Chen}, \bibinfo{person}{Rui Zhong},
  \bibinfo{person}{Yupeng Yang}, \bibinfo{person}{Hong Hu},
  \bibinfo{person}{Dinghao Wu}, {and} \bibinfo{person}{Wenke Lee}.}
  \bibinfo{year}{2023}\natexlab{}.
\newblock \showarticletitle{{{\textmu}FUZZ}: Redesign of Parallel Fuzzing using
  Microservice Architecture}. In \bibinfo{booktitle}{\emph{32nd USENIX Security
  Symposium (USENIX Security 23)}}. \bibinfo{publisher}{USENIX Association},
  \bibinfo{address}{Anaheim, CA}, \bibinfo{pages}{1325--1342}.
\newblock
\showISBNx{978-1-939133-37-3}
\urldef\tempurl%
\url{https://www.usenix.org/conference/usenixsecurity23/presentation/chen-yongheng}
\showURL{%
\tempurl}


\bibitem[Clause et~al\mbox{.}(2007)]%
        {dytan02}
\bibfield{author}{\bibinfo{person}{James Clause}, \bibinfo{person}{Wanchun Li},
  {and} \bibinfo{person}{Alessandro Orso}.} \bibinfo{year}{2007}\natexlab{}.
\newblock \showarticletitle{Dytan: A Generic Dynamic Taint Analysis Framework}.
  In \bibinfo{booktitle}{\emph{Proceedings of the 2007 International Symposium
  on Software Testing and Analysis}} (London, United Kingdom)
  \emph{(\bibinfo{series}{ISSTA '07})}. \bibinfo{publisher}{Association for
  Computing Machinery}, \bibinfo{address}{New York, NY, USA},
  \bibinfo{pages}{196–206}.
\newblock
\showISBNx{9781595937346}
\urldef\tempurl%
\url{https://doi.org/10.1145/1273463.1273490}
\showDOI{\tempurl}


\bibitem[Deng et~al\mbox{.}(2023a)]%
        {deng2023titan}
\bibfield{author}{\bibinfo{person}{Yinlin Deng},
  \bibinfo{person}{Chunqiu~Steven Xia}, \bibinfo{person}{Haoran Peng},
  \bibinfo{person}{Chenyuan Yang}, {and} \bibinfo{person}{Lingming Zhang}.}
  \bibinfo{year}{2023}\natexlab{a}.
\newblock \bibinfo{title}{Large Language Models are Zero-Shot Fuzzers: Fuzzing
  Deep-Learning Libraries via Large Language Models}.
\newblock
\newblock
\showeprint[arxiv]{2212.14834}~[cs.SE]


\bibitem[Deng et~al\mbox{.}(2023b)]%
        {deng2023fuzzgpt}
\bibfield{author}{\bibinfo{person}{Yinlin Deng},
  \bibinfo{person}{Chunqiu~Steven Xia}, \bibinfo{person}{Chenyuan Yang},
  \bibinfo{person}{Shizhuo~Dylan Zhang}, \bibinfo{person}{Shujing Yang}, {and}
  \bibinfo{person}{Lingming Zhang}.} \bibinfo{year}{2023}\natexlab{b}.
\newblock \bibinfo{title}{Large Language Models are Edge-Case Fuzzers: Testing
  Deep Learning Libraries via FuzzGPT}.
\newblock
\newblock
\showeprint[arxiv]{2304.02014}~[cs.SE]


\bibitem[Fioraldi et~al\mbox{.}(2020)]%
        {AFLplusplus-Woot20}
\bibfield{author}{\bibinfo{person}{Andrea Fioraldi}, \bibinfo{person}{Dominik
  Maier}, \bibinfo{person}{Heiko Ei{\ss}feldt}, {and} \bibinfo{person}{Marc
  Heuse}.} \bibinfo{year}{2020}\natexlab{}.
\newblock \showarticletitle{{AFL++}: Combining Incremental Steps of Fuzzing
  Research}. In \bibinfo{booktitle}{\emph{14th {USENIX} Workshop on Offensive
  Technologies ({WOOT} 20)}}. \bibinfo{publisher}{{USENIX} Association}.
\newblock


\bibitem[Fu et~al\mbox{.}(2017)]%
        {geabase}
\bibfield{author}{\bibinfo{person}{Zhisong Fu}, \bibinfo{person}{Zhengwei Wu},
  \bibinfo{person}{Houyi Li}, \bibinfo{person}{Yize Li}, \bibinfo{person}{Min
  Wu}, \bibinfo{person}{Xiaojie Chen}, \bibinfo{person}{Xiaomeng Ye},
  \bibinfo{person}{Benquan Yu}, {and} \bibinfo{person}{Xi Hu}.}
  \bibinfo{year}{2017}\natexlab{}.
\newblock \showarticletitle{GeaBase: A High-Performance Distributed Graph
  Database for Industry-Scale Applications}. In \bibinfo{booktitle}{\emph{2017
  Fifth International Conference on Advanced Cloud and Big Data (CBD)}}.
  \bibinfo{pages}{170--175}.
\newblock
\urldef\tempurl%
\url{https://doi.org/10.1109/CBD.2017.37}
\showDOI{\tempurl}


\bibitem[Gan et~al\mbox{.}(2018)]%
        {collAFL}
\bibfield{author}{\bibinfo{person}{Shuitao Gan}, \bibinfo{person}{Chao Zhang},
  \bibinfo{person}{Xiaojun Qin}, \bibinfo{person}{Xuwen Tu},
  \bibinfo{person}{Kang Li}, \bibinfo{person}{Zhongyu Pei}, {and}
  \bibinfo{person}{Zuoning Chen}.} \bibinfo{year}{2018}\natexlab{}.
\newblock \showarticletitle{CollAFL: Path Sensitive Fuzzing}. In
  \bibinfo{booktitle}{\emph{2018 IEEE Symposium on Security and Privacy (SP)}}.
  \bibinfo{pages}{679--696}.
\newblock
\urldef\tempurl%
\url{https://doi.org/10.1109/SP.2018.00040}
\showDOI{\tempurl}


\bibitem[Godefroid et~al\mbox{.}(2017)]%
        {rnn2}
\bibfield{author}{\bibinfo{person}{Patrice Godefroid}, \bibinfo{person}{Hila
  Peleg}, {and} \bibinfo{person}{Rishabh Singh}.}
  \bibinfo{year}{2017}\natexlab{}.
\newblock \showarticletitle{Learn{\&}Fuzz: Machine Learning for Input Fuzzing}.
\newblock \bibinfo{journal}{\emph{CoRR}}  \bibinfo{volume}{abs/1701.07232}
  (\bibinfo{year}{2017}).
\newblock
\showeprint[arXiv]{1701.07232}
\urldef\tempurl%
\url{http://arxiv.org/abs/1701.07232}
\showURL{%
\tempurl}


\bibitem[Google.(2018)]%
        {honggfuzzer}
\bibfield{author}{\bibinfo{person}{Google.}} \bibinfo{year}{2018}\natexlab{}.
\newblock \bibinfo{booktitle}{\emph{honggfuzz}}.
\newblock
\urldef\tempurl%
\url{hhttps://github.com/google/honggfuzz}
\showURL{%
\tempurl}


\bibitem[Google(2022)]%
        {googleData}
\bibfield{author}{\bibinfo{person}{Google}.} \bibinfo{year}{2022}\natexlab{}.
\newblock \bibinfo{booktitle}{\emph{ClusterFuzz Trophies}}.
\newblock
\urldef\tempurl%
\url{https://google.github.io/clusterfuzz#trophies}
\showURL{%
\tempurl}


\bibitem[Group(2020)]%
        {drm}
\bibfield{author}{\bibinfo{person}{Ant Group}.}
  \bibinfo{year}{2020}\natexlab{}.
\newblock \bibinfo{booktitle}{\emph{Introduction to SOFAStack Microservices)}}.
\newblock


\bibitem[Group(2021a)]%
        {antscheduler}
\bibfield{author}{\bibinfo{person}{Ant Group}.}
  \bibinfo{year}{2021}\natexlab{a}.
\newblock \bibinfo{booktitle}{\emph{AntScheduler}}.
\newblock
\urldef\tempurl%
\url{https://github.com/mcalus3/AntScheduler}
\showURL{%
\tempurl}


\bibitem[Group(2021b)]%
        {sofamq}
\bibfield{author}{\bibinfo{person}{Ant Group}.}
  \bibinfo{year}{2021}\natexlab{b}.
\newblock \bibinfo{booktitle}{\emph{SOFAMQ}}.
\newblock
\urldef\tempurl%
\url{https://github.com/sofastack-guides/sofamq-demo}
\showURL{%
\tempurl}


\bibitem[Group(2021c)]%
        {sofa}
\bibfield{author}{\bibinfo{person}{Ant Group}.}
  \bibinfo{year}{2021}\natexlab{c}.
\newblock \bibinfo{booktitle}{\emph{SOFASTACK}}.
\newblock
\urldef\tempurl%
\url{https://github.com/sofastack}
\showURL{%
\tempurl}


\bibitem[G\"{u}ler et~al\mbox{.}(2020)]%
        {Cupid}
\bibfield{author}{\bibinfo{person}{Emre G\"{u}ler}, \bibinfo{person}{Philipp
  G\"{o}rz}, \bibinfo{person}{Elia Geretto}, \bibinfo{person}{Andrea Jemmett},
  \bibinfo{person}{Sebastian \"{O}sterlund}, \bibinfo{person}{Herbert Bos},
  \bibinfo{person}{Cristiano Giuffrida}, {and} \bibinfo{person}{Thorsten
  Holz}.} \bibinfo{year}{2020}\natexlab{}.
\newblock \showarticletitle{Cupid: Automatic Fuzzer Selection for Collaborative
  Fuzzing}. In \bibinfo{booktitle}{\emph{Annual Computer Security Applications
  Conference}} (Austin, USA) \emph{(\bibinfo{series}{ACSAC '20})}.
  \bibinfo{publisher}{Association for Computing Machinery},
  \bibinfo{address}{New York, NY, USA}, \bibinfo{pages}{360–372}.
\newblock
\showISBNx{9781450388580}
\urldef\tempurl%
\url{https://doi.org/10.1145/3427228.3427266}
\showDOI{\tempurl}


\bibitem[Lemieux and Sen(2018)]%
        {fairfuzz}
\bibfield{author}{\bibinfo{person}{Caroline Lemieux} {and}
  \bibinfo{person}{Koushik Sen}.} \bibinfo{year}{2018}\natexlab{}.
\newblock \showarticletitle{FairFuzz: A Targeted Mutation Strategy for
  Increasing Greybox Fuzz Testing Coverage}. In \bibinfo{booktitle}{\emph{2018
  33rd IEEE/ACM International Conference on Automated Software Engineering
  (ASE)}}. \bibinfo{pages}{475--485}.
\newblock
\urldef\tempurl%
\url{https://doi.org/10.1145/3238147.3238176}
\showDOI{\tempurl}


\bibitem[Leung and White(1989)]%
        {regression01}
\bibfield{author}{\bibinfo{person}{Hareton K.~N. Leung} {and}
  \bibinfo{person}{Lee~J. White}.} \bibinfo{year}{1989}\natexlab{}.
\newblock \showarticletitle{Insights into regression testing (software
  testing)}.
\newblock \bibinfo{journal}{\emph{Proceedings. Conference on Software
  Maintenance - 1989}} (\bibinfo{year}{1989}), \bibinfo{pages}{60--69}.
\newblock


\bibitem[Li et~al\mbox{.}(2017)]%
        {steer}
\bibfield{author}{\bibinfo{person}{Yuekang Li}, \bibinfo{person}{Bihuan Chen},
  \bibinfo{person}{Mahinthan Chandramohan}, \bibinfo{person}{Shang-Wei Lin},
  \bibinfo{person}{Yang Liu}, {and} \bibinfo{person}{Alwen Tiu}.}
  \bibinfo{year}{2017}\natexlab{}.
\newblock \showarticletitle{Steelix: Program-State Based Binary Fuzzing}. In
  \bibinfo{booktitle}{\emph{Proceedings of the 2017 11th Joint Meeting on
  Foundations of Software Engineering}} (Paderborn, Germany)
  \emph{(\bibinfo{series}{ESEC/FSE 2017})}. \bibinfo{publisher}{Association for
  Computing Machinery}, \bibinfo{address}{New York, NY, USA},
  \bibinfo{pages}{627–637}.
\newblock
\showISBNx{9781450351058}
\urldef\tempurl%
\url{https://doi.org/10.1145/3106237.3106295}
\showDOI{\tempurl}


\bibitem[Liang et~al\mbox{.}(2018)]%
        {PAFL}
\bibfield{author}{\bibinfo{person}{Jie Liang}, \bibinfo{person}{Yu Jiang},
  \bibinfo{person}{Yuanliang Chen}, \bibinfo{person}{Mingzhe Wang},
  \bibinfo{person}{Chijin Zhou}, {and} \bibinfo{person}{Jiaguang Sun}.}
  \bibinfo{year}{2018}\natexlab{}.
\newblock \showarticletitle{PAFL: Extend Fuzzing Optimizations of Single Mode
  to Industrial Parallel Mode}. In \bibinfo{booktitle}{\emph{Proceedings of the
  2018 26th ACM Joint Meeting on European Software Engineering Conference and
  Symposium on the Foundations of Software Engineering}} (Lake Buena Vista, FL,
  USA) \emph{(\bibinfo{series}{ESEC/FSE 2018})}.
  \bibinfo{publisher}{Association for Computing Machinery},
  \bibinfo{address}{New York, NY, USA}, \bibinfo{pages}{809–814}.
\newblock
\showISBNx{9781450355735}
\urldef\tempurl%
\url{https://doi.org/10.1145/3236024.3275525}
\showDOI{\tempurl}


\bibitem[Liu et~al\mbox{.}(2022)]%
        {rockcode}
\bibfield{author}{\bibinfo{person}{Jiangchao Liu}, \bibinfo{person}{Jierui
  Liu}, \bibinfo{person}{Peng Di}, \bibinfo{person}{Alex~X. Liu}, {and}
  \bibinfo{person}{Zexin Zhong}.} \bibinfo{year}{2022}\natexlab{}.
\newblock \showarticletitle{Record and Replay of Online Traffic for
  Microservices with Automatic Mocking Point Identification}. In
  \bibinfo{booktitle}{\emph{Proceedings of the 44th International Conference on
  Software Engineering: Software Engineering in Practice}} (Pittsburgh,
  Pennsylvania) \emph{(\bibinfo{series}{ICSE-SEIP '22})}.
  \bibinfo{publisher}{Association for Computing Machinery},
  \bibinfo{address}{New York, NY, USA}, \bibinfo{pages}{221–230}.
\newblock
\showISBNx{9781450392266}
\urldef\tempurl%
\url{https://doi.org/10.1145/3510457.3513029}
\showDOI{\tempurl}


\bibitem[Liu et~al\mbox{.}(2023)]%
        {10.1145/3551349.3556929}
\bibfield{author}{\bibinfo{person}{Zixi Liu}, \bibinfo{person}{Yang Feng},
  \bibinfo{person}{Yining Yin}, \bibinfo{person}{Jingyu Sun},
  \bibinfo{person}{Zhenyu Chen}, {and} \bibinfo{person}{Baowen Xu}.}
  \bibinfo{year}{2023}\natexlab{}.
\newblock \showarticletitle{QATest: A Uniform Fuzzing Framework for Question
  Answering Systems}. In \bibinfo{booktitle}{\emph{Proceedings of the 37th
  IEEE/ACM International Conference on Automated Software Engineering}}
  (Rochester, MI, USA) \emph{(\bibinfo{series}{ASE '22})}.
  \bibinfo{publisher}{Association for Computing Machinery},
  \bibinfo{address}{New York, NY, USA}, Article \bibinfo{articleno}{81},
  \bibinfo{numpages}{12}~pages.
\newblock
\showISBNx{9781450394758}
\urldef\tempurl%
\url{https://doi.org/10.1145/3551349.3556929}
\showDOI{\tempurl}


\bibitem[LLVM(2018)]%
        {libFuzzer}
\bibfield{author}{\bibinfo{person}{LLVM}.} \bibinfo{year}{2018}\natexlab{}.
\newblock \bibinfo{booktitle}{\emph{libFuzzer}}.
\newblock
\urldef\tempurl%
\url{https://llvm.org/docs/LibFuzzer.html}
\showURL{%
\tempurl}


\bibitem[Long et~al\mbox{.}(2020)]%
        {rts03}
\bibfield{author}{\bibinfo{person}{Zhenyue Long}, \bibinfo{person}{Zeliu Ao},
  \bibinfo{person}{Guoquan Wu}, \bibinfo{person}{Wei Chen}, {and}
  \bibinfo{person}{Jun Wei}.} \bibinfo{year}{2020}\natexlab{}.
\newblock \showarticletitle{WebRTS: A Dynamic Regression Test Selection Tool
  for Java Web Applications}. In \bibinfo{booktitle}{\emph{2020 IEEE
  International Conference on Software Maintenance and Evolution (ICSME)}}.
  \bibinfo{pages}{822--825}.
\newblock
\urldef\tempurl%
\url{https://doi.org/10.1109/ICSME46990.2020.00102}
\showDOI{\tempurl}


\bibitem[Ltd({[n.\,d.]})]%
        {redis}
\bibfield{author}{\bibinfo{person}{Redis Ltd}.}
  \bibinfo{year}{[n.\,d.]}\natexlab{}.
\newblock \bibinfo{booktitle}{\emph{Redis}}.
\newblock
\urldef\tempurl%
\url{https://redis.io/}
\showURL{%
\tempurl}


\bibitem[Microsoft(2022)]%
        {aside}
\bibfield{author}{\bibinfo{person}{Microsoft}.}
  \bibinfo{year}{2022}\natexlab{}.
\newblock \bibinfo{booktitle}{\emph{Cache-Aside pattern}}.
\newblock
\urldef\tempurl%
\url{https://learn.microsoft.com/en-us/azure/architecture/patterns/cache-aside}
\showURL{%
\tempurl}


\bibitem[Newman(2021)]%
        {newman2021building}
\bibfield{author}{\bibinfo{person}{Sam Newman}.}
  \bibinfo{year}{2021}\natexlab{}.
\newblock \bibinfo{booktitle}{\emph{Building microservices}}.
\newblock \bibinfo{publisher}{" O'Reilly Media, Inc."}.
\newblock


\bibitem[OpenTracing({[n.\,d.]})]%
        {opentrace}
\bibfield{author}{\bibinfo{person}{OpenTracing}.}
  \bibinfo{year}{[n.\,d.]}\natexlab{}.
\newblock \bibinfo{booktitle}{}.
\newblock
\urldef\tempurl%
\url{https://opentracing.io/}
\showURL{%
\tempurl}


\bibitem[Peng et~al\mbox{.}(2019)]%
        {peng20191dvul}
\bibfield{author}{\bibinfo{person}{Jiaqi Peng}, \bibinfo{person}{Feng Li},
  \bibinfo{person}{Bingchang Liu}, \bibinfo{person}{Lili Xu},
  \bibinfo{person}{Binghong Liu}, \bibinfo{person}{Kai Chen}, {and}
  \bibinfo{person}{Wei Huo}.} \bibinfo{year}{2019}\natexlab{}.
\newblock \showarticletitle{1dvul: Discovering 1-day vulnerabilities through
  binary patches}. In \bibinfo{booktitle}{\emph{2019 49th Annual IEEE/IFIP
  International Conference on Dependable Systems and Networks (DSN)}}. IEEE,
  \bibinfo{pages}{605--616}.
\newblock


\bibitem[Petsios et~al\mbox{.}(2017)]%
        {NEZHA}
\bibfield{author}{\bibinfo{person}{Theofilos Petsios}, \bibinfo{person}{Adrian
  Tang}, \bibinfo{person}{Salvatore Stolfo}, \bibinfo{person}{Angelos~D.
  Keromytis}, {and} \bibinfo{person}{Suman Jana}.}
  \bibinfo{year}{2017}\natexlab{}.
\newblock \showarticletitle{NEZHA: Efficient Domain-Independent Differential
  Testing}. In \bibinfo{booktitle}{\emph{2017 IEEE Symposium on Security and
  Privacy (SP)}}. \bibinfo{pages}{615--632}.
\newblock
\urldef\tempurl%
\url{https://doi.org/10.1109/SP.2017.27}
\showDOI{\tempurl}


\bibitem[Pham et~al\mbox{.}(2021)]%
        {Pham2021}
\bibfield{author}{\bibinfo{person}{Van-Thuan Pham}, \bibinfo{person}{Manh-Dung
  Nguyen}, \bibinfo{person}{Quang-Trung Ta}, \bibinfo{person}{Toby Murray},
  {and} \bibinfo{person}{Benjamin~I.P. Rubinstein}.}
  \bibinfo{year}{2021}\natexlab{}.
\newblock \showarticletitle{Towards Systematic and Dynamic Task Allocation for
  Collaborative Parallel Fuzzing}. In \bibinfo{booktitle}{\emph{2021 36th
  IEEE/ACM International Conference on Automated Software Engineering (ASE)}}.
  \bibinfo{pages}{1337--1341}.
\newblock
\urldef\tempurl%
\url{https://doi.org/10.1109/ASE51524.2021.9678810}
\showDOI{\tempurl}


\bibitem[Rajpal et~al\mbox{.}(2017)]%
        {lstm}
\bibfield{author}{\bibinfo{person}{Mohit Rajpal}, \bibinfo{person}{William
  Blum}, {and} \bibinfo{person}{Rishabh Singh}.}
  \bibinfo{year}{2017}\natexlab{}.
\newblock \showarticletitle{Not all bytes are equal: Neural byte sieve for
  fuzzing}.
\newblock \bibinfo{journal}{\emph{CoRR}}  \bibinfo{volume}{abs/1711.04596}
  (\bibinfo{year}{2017}).
\newblock
\showeprint[arXiv]{1711.04596}
\urldef\tempurl%
\url{http://arxiv.org/abs/1711.04596}
\showURL{%
\tempurl}


\bibitem[She et~al\mbox{.}(2019)]%
        {neuzz}
\bibfield{author}{\bibinfo{person}{Dongdong She}, \bibinfo{person}{Kexin Pei},
  \bibinfo{person}{Dave Epstein}, \bibinfo{person}{Junfeng Yang},
  \bibinfo{person}{Baishakhi Ray}, {and} \bibinfo{person}{Suman Jana}.}
  \bibinfo{year}{2019}\natexlab{}.
\newblock \showarticletitle{NEUZZ: Efficient Fuzzing with Neural Program
  Smoothing}. In \bibinfo{booktitle}{\emph{2019 IEEE Symposium on Security and
  Privacy (SP)}}. \bibinfo{pages}{803--817}.
\newblock
\urldef\tempurl%
\url{https://doi.org/10.1109/SP.2019.00052}
\showDOI{\tempurl}


\bibitem[Sridharan et~al\mbox{.}(2011)]%
        {taintr02}
\bibfield{author}{\bibinfo{person}{Manu Sridharan}, \bibinfo{person}{Shay
  Artzi}, \bibinfo{person}{Marco Pistoia}, \bibinfo{person}{Salvatore
  Guarnieri}, \bibinfo{person}{Omer Tripp}, {and} \bibinfo{person}{Ryan Berg}.}
  \bibinfo{year}{2011}\natexlab{}.
\newblock \showarticletitle{F4F: Taint Analysis of Framework-Based Web
  Applications}. In \bibinfo{booktitle}{\emph{Proceedings of the 2011 ACM
  International Conference on Object Oriented Programming Systems Languages and
  Applications}} (Portland, Oregon, USA) \emph{(\bibinfo{series}{OOPSLA '11})}.
  \bibinfo{publisher}{Association for Computing Machinery},
  \bibinfo{address}{New York, NY, USA}, \bibinfo{pages}{1053–1068}.
\newblock
\showISBNx{9781450309400}
\urldef\tempurl%
\url{https://doi.org/10.1145/2048066.2048145}
\showDOI{\tempurl}


\bibitem[Wang et~al\mbox{.}(2023)]%
        {10298503}
\bibfield{author}{\bibinfo{person}{W. Wang}, \bibinfo{person}{A. Benea}, {and}
  \bibinfo{person}{F. Ivancic}.} \bibinfo{year}{2023}\natexlab{}.
\newblock \showarticletitle{Zero-Config Fuzzing for Microservices}. In
  \bibinfo{booktitle}{\emph{2023 38th IEEE/ACM International Conference on
  Automated Software Engineering (ASE)}}. \bibinfo{publisher}{IEEE Computer
  Society}, \bibinfo{address}{Los Alamitos, CA, USA},
  \bibinfo{pages}{1840--1845}.
\newblock
\urldef\tempurl%
\url{https://doi.org/10.1109/ASE56229.2023.00036}
\showDOI{\tempurl}


\bibitem[Wang et~al\mbox{.}(2021)]%
        {wang2021}
\bibfield{author}{\bibinfo{person}{Yifan Wang}, \bibinfo{person}{Yuchen Zhang},
  \bibinfo{person}{Chenbin Pang}, \bibinfo{person}{Peng Li},
  \bibinfo{person}{Nikolaos Triandopoulos}, {and} \bibinfo{person}{Jun Xu}.}
  \bibinfo{year}{2021}\natexlab{}.
\newblock \showarticletitle{{Facilitating Parallel Fuzzing with
  Mutually-Exclusive Task Distribution}}. In \bibinfo{booktitle}{\emph{Secur.
  Priv. Commun. Networks}}, \bibfield{editor}{\bibinfo{person}{Joaquin
  Garcia-Alfaro}, \bibinfo{person}{Shujun Li}, \bibinfo{person}{Radha
  Poovendran}, \bibinfo{person}{Herv{\'{e}} Debar}, {and} \bibinfo{person}{Moti
  Yung}} (Eds.). \bibinfo{publisher}{Springer International Publishing},
  \bibinfo{address}{Cham}, \bibinfo{pages}{185--206}.
\newblock
\showISBNx{978-3-030-90022-9}


\bibitem[Wiki({[n.\,d.]})]%
        {dijkstra}
\bibfield{author}{\bibinfo{person}{Wiki}.} \bibinfo{year}{[n.\,d.]}\natexlab{}.
\newblock \bibinfo{booktitle}{\emph{Dijkstra's Algorithm}}.
\newblock
\urldef\tempurl%
\url{https://en.wikipedia.org/wiki/Dijkstra%27s_algorithm}
\showURL{%
\tempurl}


\bibitem[Wikipedia({[n.\,d.]})]%
        {microserviceswiki}
\bibfield{author}{\bibinfo{person}{Wikipedia}.}
  \bibinfo{year}{[n.\,d.]}\natexlab{}.
\newblock \bibinfo{booktitle}{\emph{Microservices.}}
\newblock
\urldef\tempurl%
\url{https://en.wikipedia.org/wiki/Microservices}
\showURL{%
Retrieved Feb 14, 2023 from \tempurl}


\bibitem[Xu et~al\mbox{.}(2017)]%
        {Xu2017}
\bibfield{author}{\bibinfo{person}{Wen Xu}, \bibinfo{person}{Sanidhya Kashyap},
  \bibinfo{person}{Changwoo Min}, {and} \bibinfo{person}{Taesoo Kim}.}
  \bibinfo{year}{2017}\natexlab{}.
\newblock \showarticletitle{Designing New Operating Primitives to Improve
  Fuzzing Performance}. In \bibinfo{booktitle}{\emph{Proceedings of the 2017
  ACM SIGSAC Conference on Computer and Communications Security}} (Dallas,
  Texas, USA) \emph{(\bibinfo{series}{CCS '17})}.
  \bibinfo{publisher}{Association for Computing Machinery},
  \bibinfo{address}{New York, NY, USA}, \bibinfo{pages}{2313–2328}.
\newblock
\showISBNx{9781450349468}
\urldef\tempurl%
\url{https://doi.org/10.1145/3133956.3134046}
\showDOI{\tempurl}


\bibitem[Yoo and Harman(2012)]%
        {regress_survey}
\bibfield{author}{\bibinfo{person}{S. Yoo} {and} \bibinfo{person}{M. Harman}.}
  \bibinfo{year}{2012}\natexlab{}.
\newblock \showarticletitle{Regression Testing Minimization, Selection and
  Prioritization: A Survey}.
\newblock \bibinfo{journal}{\emph{Softw. Test. Verif. Reliab.}}
  \bibinfo{volume}{22}, \bibinfo{number}{2} (\bibinfo{date}{mar}
  \bibinfo{year}{2012}), \bibinfo{pages}{67–120}.
\newblock
\showISSN{0960-0833}
\urldef\tempurl%
\url{https://doi.org/10.1002/stv.430}
\showDOI{\tempurl}


\bibitem[Zalewski.(2014)]%
        {afl}
\bibfield{author}{\bibinfo{person}{Michal Zalewski.}}
  \bibinfo{year}{2014}\natexlab{}.
\newblock \bibinfo{booktitle}{\emph{American Fuzzing Loop}}.
\newblock
\urldef\tempurl%
\url{https://lcamtuf.coredump.cx/afl/}
\showURL{%
\tempurl}


\bibitem[Zhang(2018)]%
        {rts02}
\bibfield{author}{\bibinfo{person}{Lingming Zhang}.}
  \bibinfo{year}{2018}\natexlab{}.
\newblock \showarticletitle{Hybrid Regression Test Selection}. In
  \bibinfo{booktitle}{\emph{Proceedings of the 40th International Conference on
  Software Engineering}} (Gothenburg, Sweden) \emph{(\bibinfo{series}{ICSE
  '18})}. \bibinfo{publisher}{Association for Computing Machinery},
  \bibinfo{address}{New York, NY, USA}, \bibinfo{pages}{199–209}.
\newblock
\showISBNx{9781450356381}
\urldef\tempurl%
\url{https://doi.org/10.1145/3180155.3180198}
\showDOI{\tempurl}


\bibitem[Zhang et~al\mbox{.}(2023)]%
        {10.1145/3585009}
\bibfield{author}{\bibinfo{person}{Man Zhang}, \bibinfo{person}{Andrea Arcuri},
  \bibinfo{person}{Yonggang Li}, \bibinfo{person}{Yang Liu}, {and}
  \bibinfo{person}{Kaiming Xue}.} \bibinfo{year}{2023}\natexlab{}.
\newblock \showarticletitle{White-Box Fuzzing RPC-Based APIs with EvoMaster: An
  Industrial Case Study}.
\newblock \bibinfo{journal}{\emph{ACM Trans. Softw. Eng. Methodol.}}
  \bibinfo{volume}{32}, \bibinfo{number}{5}, Article \bibinfo{articleno}{122}
  (\bibinfo{date}{jul} \bibinfo{year}{2023}), \bibinfo{numpages}{38}~pages.
\newblock
\showISSN{1049-331X}
\urldef\tempurl%
\url{https://doi.org/10.1145/3585009}
\showDOI{\tempurl}


\bibitem[Zhang et~al\mbox{.}(2022)]%
        {zhang2022fuzzing}
\bibfield{author}{\bibinfo{person}{Man Zhang}, \bibinfo{person}{Andrea Arcuri},
  \bibinfo{person}{Yonggang Li}, \bibinfo{person}{Kaiming Xue},
  \bibinfo{person}{Zhao Wang}, \bibinfo{person}{Jian Huo}, {and}
  \bibinfo{person}{Weiwei Huang}.} \bibinfo{year}{2022}\natexlab{}.
\newblock \bibinfo{title}{Fuzzing Microservices In Industry: Experience of
  Applying EvoMaster at Meituan}.
\newblock
\newblock
\showeprint[arxiv]{2208.03988}~[cs.SE]


\bibitem[Zhao et~al\mbox{.}(2023)]%
        {zhao2023incremental}
\bibfield{author}{\bibinfo{person}{Zelin Zhao}, \bibinfo{person}{Xizao Wang},
  \bibinfo{person}{Zhaogui Xu}, \bibinfo{person}{Zhenhao Tang},
  \bibinfo{person}{Yongchao Li}, {and} \bibinfo{person}{Peng Di}.}
  \bibinfo{year}{2023}\natexlab{}.
\newblock \showarticletitle{Incremental Call Graph Construction in Industrial
  Practice}. In \bibinfo{booktitle}{\emph{2023 IEEE/ACM 45th International
  Conference on Software Engineering: Software Engineering in Practice
  (ICSE-SEIP)}}. IEEE, \bibinfo{pages}{471--482}.
\newblock


\bibitem[Zhong et~al\mbox{.}(2019)]%
        {rst011}
\bibfield{author}{\bibinfo{person}{Hua Zhong}, \bibinfo{person}{Lingming
  Zhang}, {and} \bibinfo{person}{Sarfraz Khurshid}.}
  \bibinfo{year}{2019}\natexlab{}.
\newblock \showarticletitle{TestSage: Regression Test Selection for Large-Scale
  Web Service Testing}. In \bibinfo{booktitle}{\emph{2019 12th IEEE Conference
  on Software Testing, Validation and Verification (ICST)}}.
  \bibinfo{pages}{430--440}.
\newblock
\urldef\tempurl%
\url{https://doi.org/10.1109/ICST.2019.00052}
\showDOI{\tempurl}


\bibitem[Zhong et~al\mbox{.}(2023)]%
        {zhong2023}
\bibfield{author}{\bibinfo{person}{Zexin Zhong}, \bibinfo{person}{Jiangchao
  Liu}, \bibinfo{person}{Diyu Wu}, \bibinfo{person}{Peng Di},
  \bibinfo{person}{Yulei Sui}, \bibinfo{person}{Alex~X. Liu}, {and}
  \bibinfo{person}{John C.~S. Lui}.} \bibinfo{year}{2023}\natexlab{}.
\newblock \showarticletitle{Scalable Compositional Static Taint Analysis for
  Sensitive Data Tracing on Industrial Micro-Services}. In
  \bibinfo{booktitle}{\emph{2023 IEEE/ACM 45th International Conference on
  Software Engineering: Software Engineering in Practice (ICSE-SEIP)}}.
  \bibinfo{pages}{110--121}.
\newblock
\urldef\tempurl%
\url{https://doi.org/10.1109/ICSE-SEIP58684.2023.00015}
\showDOI{\tempurl}


\bibitem[Zhou et~al\mbox{.}(2020)]%
        {UniFuzz}
\bibfield{author}{\bibinfo{person}{Xu Zhou}, \bibinfo{person}{Pengfei Wang},
  \bibinfo{person}{Chenyifan Liu}, \bibinfo{person}{Tai Yue},
  \bibinfo{person}{Yingying Liu}, \bibinfo{person}{Congxi Song},
  \bibinfo{person}{Kai Lu}, {and} \bibinfo{person}{Qidi Yin}.}
  \bibinfo{year}{2020}\natexlab{}.
\newblock \showarticletitle{UniFuzz: Optimizing Distributed Fuzzing via Dynamic
  Centralized Task Scheduling}.
\newblock \bibinfo{journal}{\emph{ArXiv}}  \bibinfo{volume}{abs/2009.06124}
  (\bibinfo{year}{2020}).
\newblock
\urldef\tempurl%
\url{https://api.semanticscholar.org/CorpusID:221655823}
\showURL{%
\tempurl}


\bibitem[Zong et~al\mbox{.}(2020)]%
        {fuzzguard}
\bibfield{author}{\bibinfo{person}{Peiyuan Zong}, \bibinfo{person}{Tao Lv},
  \bibinfo{person}{Dawei Wang}, \bibinfo{person}{Zizhuang Deng},
  \bibinfo{person}{Ruigang Liang}, {and} \bibinfo{person}{Kai Chen}.}
  \bibinfo{year}{2020}\natexlab{}.
\newblock \showarticletitle{{FuzzGuard}: Filtering out Unreachable Inputs in
  Directed Grey-box Fuzzing through Deep Learning}. In
  \bibinfo{booktitle}{\emph{29th USENIX Security Symposium (USENIX Security
  20)}}. \bibinfo{publisher}{USENIX Association}, \bibinfo{pages}{2255--2269}.
\newblock
\showISBNx{978-1-939133-17-5}
\urldef\tempurl%
\url{https://www.usenix.org/conference/usenixsecurity20/presentation/zong}
\showURL{%
\tempurl}


\end{thebibliography}
\end{document}